\documentclass[aps,prb,twocolumn,superscriptaddress,showpacs]{revtex4}
\usepackage{amsmath}
\usepackage{amsfonts}
\usepackage{amssymb}
\usepackage{amsthm}
\usepackage[final]{graphicx}
\usepackage[latin1]{inputenc}
\usepackage[T1]{fontenc}
\usepackage{lmodern}
\usepackage{color}
\usepackage{dsfont}

\newcommand{\D}{\text{D}}

\newcommand{\Ew}[1]{\langle #1\rangle}
\newcommand{\kompm}[2]{[#1,#2]_\pm}
\newcommand{\kom}[2]{[#1,#2]_-}
\newcommand{\s}{\sigma}
\newcommand{\da}{^{\dagger}}
\newcommand{\p}{^{\prime}}
\newcommand{\di}{\mathrm{d}}
\newcommand{\dd}[2]{\frac{\di#1}{\di#2}}
\newcommand{\sgn}{\text{sgn}}
\newcommand{\Tr}{\text{Tr}}
\newcommand{\nd}{{\text{nd}}}
\newcommand{\st}{{\text{st}}}
\newcommand{\eff}{{\text{eff}}}
\newcommand{\res}{{\text{res}}}
\newcommand{\red}{{\text{red}}}

\renewcommand{\Im}{\text{Im}\,}
\renewcommand{\Re}{\text{Re}\,}
\newcommand{\II}{{I\!I}}

\newcommand{\wt}{\widetilde}
\newcommand{\akom}[2]{[#1,#2]_+}

\newcommand{\mc}[1]{\mathcal{#1}}

\newcommand{\bG}{\bar{G}}
\newcommand{\bI}{\bar{I}}
\newcommand{\bo}{\bar{\omega}}
\newcommand{\1}{\bar{1}}
\newcommand{\2}{\bar{2}}
\newcommand{\3}{\bar{3}}
\newcommand{\tr}{\text{tr}}
\newcommand{\ddl}[1]{\dd{#1}{\Lambda}}
\newcommand{\E}{\mathds{1}}
\newcommand{\wtK}{\wt{K}_\Lambda}
\newcommand{\IAB}{I\!I}
\newcommand{\bIAB}{\bar{I\!I}}
\newcommand{\bII}{\bar{I\!I}}

\def\vec#1{\mathchoice{\mbox{\boldmath$\displaystyle#1$}} 
{\mbox{\boldmath$\textstyle#1$}} 
{\mbox{\boldmath$\scriptstyle#1$}} 
{\mbox{\boldmath$\scriptscriptstyle#1$}}}

\allowdisplaybreaks

\urldef{\pacsurl}{\url}{http://www.aip.org/pacs/pacs2010/individuals/pacs2010_regular_edition/reg70.htm#72}

\begin{document}
\title{Magnetic field effects on the finite-frequency noise and ac conductance of a Kondo quantum dot out of equilibrium}
\author{Sarah Y. M\"uller}
\affiliation{Faculty of Physics, University of Vienna, Boltzmanngasse 5, 1090 Vienna, Austria}
\affiliation{Institute for Theory of Statistical Physics, RWTH Aachen University and 
JARA--Fundamentals of Future Information Technology, 52056 Aachen, Germany}
\author{Mikhail Pletyukhov}\author{Dirk Schuricht}
\affiliation{Institute for Theory of Statistical Physics, RWTH Aachen University and 
JARA--Fundamentals of Future Information Technology, 52056 Aachen, Germany}
\author{Sabine Andergassen}
\affiliation{Faculty of Physics, University of Vienna, Boltzmanngasse 5, 1090 Vienna, Austria}

\begin{abstract}
We present analytic results for the finite-frequency current noise and the nonequilibrium ac conductance for a Kondo quantum dot in presence of a magnetic field. Using the real-time renormalization group method, 
we determine the line shape close to resonances and show that while all resonances in the ac conductance are broadened by the transverse spin relaxation rate, the noise at finite field additionally involves the longitudinal rate as well as sharp kinks resulting in singular derivatives. 
Our results provide a consistent theoretical description of recent experimental data for the emission noise at zero magnetic field, and we propose the extension to finite field for which we present a detailed prediction.

\end{abstract}


\maketitle

\section{Introduction}\label{sec:introduction}

The understanding of quantum many-body effects and their characteristic signatures in transport properties represents a fundamental topic in mesoscopic physics. 
Beside the average current, its fluctuations described by the current noise contain additional information on 
the interplay of strong correlations and quantum fluctuations. In particular, the finite-frequency noise reveals the characteristic time scales of the system and provides information about the dynamics of excitations. 
The developments in the nanoscale device fabrication technology led to the experimental analysis of the noise
in various systems ranging from Josephson junctions to single-electron transistors~\cite{noise}.

It is by now well established that strong correlations play a crucial role for the transport properties of quantum dots. For example, for quantum dots in the so-called Kondo regime the transport is dominated by spin fluctuations leading, at sufficiently low energies, to a universal conductance of $G=2e^2/h$ due to resonant tunneling processes~\cite{G0}. Recently it has also become possible to measure the current noise in such Kondo quantum dots realized in carbon-nanotube devices.~\cite{Delattre,PhysRevLett.108.046802} In particular, Basset \emph{et al.}~\cite{PhysRevLett.108.046802} measured the finite-frequency emission noise and observed resonances when the external frequency equaled the applied bias voltage.

The nonequilibrium finite-frequency noise in quantum dots has theoretically been studied for the Anderson model, resonant level models, and spin valve systems~\cite{noisetheory}. For quantum dots in the Kondo regime previous studies focused either on the shot noise (zero frequency)~\cite{meir} or on the exactly solvable Toulouse limit~\cite{PhysRevB.58.14978}, while the finite-frequency noise has only very recently started to attract attention~\cite{PhysRevLett.108.046802,Korb, moca,PhysRevB.83.201303}. Of particular interest in this context is the nontrivial interplay of the different energy scales, which manifests itself in the appearance of characteristic resonances whose line shapes contain information about the underlying microscopic relaxation mechanisms. For quantum dots in the Kondo regime these are the transverse and longitudinal relaxation of the dot spin, which are identical at zero magnetic field, but acquire different values when the rotational symmetry is broken.

In this work we provide an analytic analysis of the finite-frequency current noise and the ac conductance in the nonequilibrium Kondo model. 
We apply the real-time renormalization group (RTRG) method~\cite{EPJST,PhysRevB.80.075120}, which is based on a systematic expansion in the reservoir-system coupling. Using
the solution of the two-loop RG equations we derive analytic results for the noise and ac conductance in the weak-coupling regime ${\rm max}\{|\Omega|, |V |, |h_0|\} \gg T_K$, where
$T_K$ denotes the Kondo scale at which the system enters the strong-coupling regime. 
We analyze the characteristic features in the noise and conductance in detail. We  particularly focus on the effects of a finite magnetic field and show that it leads to (i) characteristic resonances as a function of the frequency and bias voltage, and (ii) the appearance of both the longitudinal and transverse spin relaxation rates in the broadening of these resonances as well as sharp kinks in the noise. We find excellent agreement with existing experimental data~\cite{PhysRevLett.108.046802} for the emission noise at zero magnetic field, and propose the measurement at finite field for which we present a detailed analysis.

The paper is organized as follows. In the next section we define the symmetric and antisymmetric current noise as well as their relation to 
the ac conductance. After introducing the Kondo model we describe the calculation of the dynamical current-current correlation function 
with the RTRG method in Sec.~\ref{sec:tech}, the technical details are reported in the appendices.
In Secs.~\ref{sec:noise}, \ref{sec:exp} we present the analytic results for the finite-frequency current noise obtained from the solution of the 
flow equations and discuss their experimental observation in connection with recent data.~\cite{PhysRevLett.108.046802} 
We finally determine the real and imaginary part of the nonequilibrium ac conductance, and conclude with a summary.

\section{Current-current correlations and ac conductance}\label{sec:general}

The nonequilibrium dc current through Kondo quantum dots has been intensively studied~\cite{Kondotransport,PhysRevB.80.045117,arxiv1201.6295} in the past. Here we investigate the zero-temperature fluctuations of the current in the stationary state, which are captured
by the symmetric and antisymmetric current noise
\begin{equation}
S^\pm(t)=\frac{1}{2}\Ew{\kompm{I(t)-\Ew{I}}{I(0)-\Ew{I}}}\,,
\label{eq:def_S_t}
\end{equation}
where $I= -\dot{N}_L=-i\kom{H}{N_L}$ denotes the current operator 
in the left lead with the corresponding particle number $N_L$, and $\langle I \rangle$ is the stationary current. 
Due to the fixed number of electrons on the dot, other lead components of the noise are given by $S_{\alpha \beta}^\pm(\Omega) = \alpha\beta S^{\pm} (\Omega)$ with $\alpha,\beta = \pm$ for left/right leads.
The finite-frequency noise refers to the Fourier transform
\begin{equation}
S^\pm(\Omega)= \int_{-\infty}^\infty\di t e^{i\Omega t} S^\pm(t) \,,
\label{eq:def_S}
\end{equation}
with $S^\pm(\Omega)=\pm S^\pm(-\Omega)$.
The symmetric and antisymmetric noise 
determine 
the absorption and emission noise induced by photon absorption and emission~\cite{Schoelkopf}
\begin{equation}
S^{a/e}(\Omega)=S^+(\Omega)\pm S^-(\Omega)\,,
\label{eq:Sae}
\end{equation}
related by $S^a(-\Omega)=S^e(\Omega)$. 
Positive (negative) frequencies correspond to photon emission (absorption).
In equilibrium, $S^+ (\Omega)$ and $S^- (\Omega)$ are related by the fluctuation dissipation theorem~\cite{landau} (FDT), which at $T=0$ reads 
\begin{equation}
S^-(\Omega)={\rm sgn}(\Omega) S^+(\Omega)\,.
\label{eq:FDT}
\end{equation} 
As a consequence, the emission (absorption) noise in equilibrium vanishes for positive (negative) frequencies. 

In order to calculate $S^\pm(\Omega)$ it is useful to introduce the auxiliary current-current correlation function 
\begin{equation}
C^\pm(\Omega)=\int\limits_{-\infty}^0\di te^{-i\Omega t}\Ew{\kompm{I(0)}{I(t)}}\,,
\label{eq:def_C}
\end{equation}
which is related to the symmetric and antisymmetric noise by~\cite{PhysRevB.80.075120}
\begin{subequations}
\label{eq:S_C}
\begin{align}
&S^+ (\Omega) = \Re C^+ (\Omega)-2\pi\Ew{I}^2\delta(\Omega) ,
\label{eq:S_Ca} \\
&S^- (\Omega) = \Re C^- (\Omega) .
\label{eq:S_Cb}
\end{align} 
\end{subequations}
The calculation of $C^\pm(\Omega)$ will be addressed in the next section.

In addition to the finite-frequency noise, we study the nonequilibrium ac conductance $G(\Omega)$ 
induced by a small ac voltage modulation of the dc bias $V(t)=V+\delta V e^{-i\Omega t}$.
The real part is determined by the antisymmetric noise \cite{Safi,PhysRevB.83.201303}
\begin{equation}
\Re G(\Omega)=\frac{S^-(\Omega)}{\Omega}\,,
\label{eq:ReG_CCC}
\end{equation}
and the imaginary part can be obtained by the Kramers-Kronig relations, 
with $\Re G(\Omega)=\Re G(-\Omega)$ and $\Im G(\Omega)=-\Im G(-\Omega)$. 
Alternatively, generalizing the Kubo formula to nonequilibrium distributions allows to derive
both the real and imaginary part of $G (\Omega)$ from the auxiliary function 
$C^-(\Omega)$ by~\cite{Kubala,Safi}
\begin{equation}
G(\Omega)=\frac{1}{\Omega}\left[C^-(\Omega)-C^-(0)\right]\,,
\label{eq:G}
\end{equation} 
without resorting to the Kramers-Kronig relations.
A detailed derivation of this extension is provided in App.~\ref{app:a}.

Combining Eqs.~\eqref{eq:Sae} and \eqref{eq:ReG_CCC}, the absorption noise is determined by the emission 
noise and the real part of the ac conductance
\begin{equation}
S^a(\Omega)=S^e(\Omega)+2\Omega\ \Re G(\Omega)\,.
\end{equation}
Alternatively, measuring the emission and the absorption noise allows to extract the ac
conductance, which may represent an advantage with respect to a direct ac detection.
We note that the real part of the ac conductance relates the symmetric noise to the emission 
noise by 
\begin{equation}
S^e(\Omega)=S^+(\Omega)-\Omega \,\Re G(\Omega)\,.
\end{equation}
Hence, the emission excess noise $\Delta S^e(\Omega)$, defined as the difference between the emission noise at finite $V$ and 
$V=0$, is given by
$ \Delta S^e(\Omega)=\Delta S^+(\Omega)-\Omega\, \Re [\Delta G(\Omega)]$.
In the linear voltage regime  the ac conductance is approximately constant in $V$, implying $\Re [\Delta G\,(\Omega)]=0$; therefore
the emission excess noise coincides with the symmetric excess noise and is thus an even function of 
frequency. In the nonlinear regime, the $V$ dependence of $\Re [\Delta G(\Omega)]$ leads to an asymmetric emission excess noise.

\section{Model and RTRG method}\label{sec:tech}

\subsection{Model}\label{sec:model}

\begin{figure}
	\centering
		\includegraphics{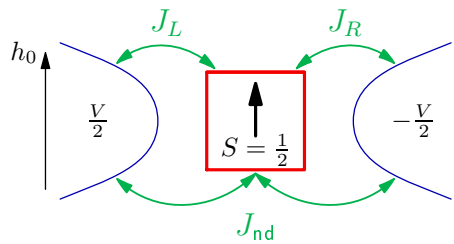}
	\caption{(Color online) Sketch of the considered quantum dot in the Kondo regime.}
	\label{fig:fig1}
\end{figure}
We consider a Kondo quantum dot consisting of a spin-1/2 $\vec{S}$ subject to a local magnetic field $h_0$, which is coupled to two noninteracting electronic leads via an isotropic exchange interaction (see Fig.~\ref{fig:fig1}),
\begin{equation}
H=H_{\rm res}+h_0S^z+
\frac{1}{2}\sum_{\alpha\alpha^\prime kk^\prime \sigma\sigma^\prime}\!\!\!\!\! J_{\alpha\alpha^\prime}
a^\dagger_{\alpha k\sigma}\vec{S}\cdot\vec{\sigma}_{\sigma\sigma^\prime}a_{\alpha' k'\sigma'}\,.
\label{eq:model}
\end{equation}
Here $a^\dagger_{\alpha k\sigma}$ and $a_{\alpha k\sigma}$ create and annihilate electrons with momentum $k$ and spin $\sigma=\ \uparrow,\downarrow$ in lead $\alpha=L,R$, and $\vec{\sigma}$ are the Pauli matrices. The leads are described by $H_\res=\sum_{\alpha k\sigma}\varepsilon_{k} a^\dagger_{\alpha k\sigma}a_{\alpha k\sigma}$, with a flat density of states in a band of width $2D$, and chemical potentials $\mu_{L/R}=\pm V/2$. The exchange interaction is assumed to be derived from an Anderson impurity model via the Schrieffer-Wolff transformation and thus satisfies $J_\nd^2=J_{L}J_{R}$, where $J_\nd=J_{RL}=J_{LR}$ and $J_\alpha=J_{\alpha\alpha}$. We use the parametrization $J_{L/R}=2x_{L/R}J_0$ with $x_L+x_R=1$. The system is at zero temperature and we use units such that $e=\hbar=k_B=2\mu_B=1$.

\subsection{RTRG method}\label{sec:method}

We calculate the current noise using the RTRG approach~\cite{EPJST,PhysRevB.80.075120}.
Here we present the essentials, for a detailed derivation including technical details we refer to App.~\ref{sec:rtrg}.

The dynamics of the reduced density matrix of the dot $\rho_\D(t)=\Tr_\res\rho(t)$, obtained by tracing out the lead degrees of freedom from the full density matrix of the system, is described by the von Neumann equation 
\begin{equation}
\dot{\rho}_\D(t)=-iL_\D\rho_\D(t)-i\int_{t_0}^t\di t'\Sigma(t-t')\rho_\D(t')\,,
\end{equation}
for an initially decoupled system $\rho(t_0)=\rho_\D(t_0)\rho_L\rho_R$ with an arbitrary dot density matrix $\rho_\D(t_0)$ and the left and right reservoir given by grand-canonical distribution functions.
The first term describes the dynamics of the isolated dot, and the dissipative kernel $\Sigma(t-t')$ 
contains the information about the effects on the local spin due to the coupling to the reservoirs. 
Introducing a Laplace variable $z$, the effective dot Liouvillian $L_\D^\eff(z)=L_\D+\Sigma(z)$
\begin{equation}
\rho_\D(z)=\int_{t_0}^\infty\di te^{iz(t-t_0)}\rho_\D(t)=\frac{i}{z-L_\D^\eff(z)}\rho_\D(t_0)\,,
\end{equation}
governs the time evolution of the reduced density matrix of the dot.
The stationary reduced density matrix is obtained by carrying out the limit $t_0\rightarrow-\infty$, or equivalently in Laplace space
\begin{equation}
\rho_\D^\st=\lim_{z\rightarrow i0^+}\frac{z}{z-L_\D^\eff(z)}\rho_\D(t_0)\,.
\end{equation}
The effective dot Liouvillian $L_\D^\eff(z)$ incorporates all information about the relaxation dynamics of the spin on the dot encoded in the renormalized magnetic field $h$ and the longitudinal and transverse spin relaxation rates $\Gamma_1$ and $\Gamma_2$.

In general, the noise \eqref{eq:def_S} is determined by the real part of the auxiliary 
function~\eqref{eq:def_C}, which can be expressed~\cite{PhysRevB.80.075120} as 
\begin{equation}
\label{eq:C}
C^\pm(\Omega)=\begin{aligned}[t]&-i\Tr_\D\left[\Sigma_I(\Omega)\frac{1}{\Omega-L_\D^\eff(\Omega)}\Sigma_I^\pm(\Omega,i0^+)\rho_\D^\st\right]\\
&-i\Tr_\D\left[\Sigma_{\II}^\pm(\Omega,i0^+)\rho_\D^\st\right]\,.
\end{aligned}
\end{equation} 
The kernels $\Sigma_{I}(\Omega)$, $\Sigma_{I}^\pm (\Omega,i0^+)$, and $\Sigma_{\II}^\pm(\Omega,i0^+)$ obey RG equations similar to that of the Liouvillian (see Appendix for further details).

The RTRG weak-coupling analysis is based on a systematic expansion in the renormalized exchange couplings around the poor man's scaling solution $J(\Lambda)$ given by $J(\Lambda)=[2 \ln(\Lambda/T_K )]^{-1}$. Here $\Lambda$ denotes the flow parameter, and the Kondo temperature is defined by $T_K = De^{-1/2J_0}$. Before $\Lambda$ reaches $T_K$ in the flow from high to low energy scales, that is in the range $\Lambda \geq \Lambda_c \gg T_K$, where $\Lambda_c = \sqrt{\Omega^2+V^2+h^2}$, we can carry out an expansion of the noise in a power series of $J(\Lambda )$. In doing so, we are able to identify which resonant features in the noise get broadened by relaxation rates and which remain sharp. The latter effect, in particular, occurs in the nonequilibrium setup at finite magnetic field and $\Omega = \pm V$. Technically, this is seen in the RTRG equations as the influence of the resolvent projection $P_0 \frac{1}{\Omega - L_\D^\eff (\Omega)}$ onto the zero eigenvalue subspace of the Liouvillian. This represents a nontrivial feature of the two-point functions \eqref{eq:def_S}, in contrast to one-point functions which receive no contribution from the zero eigenvalue subspace~\cite{EPJST}.

\section{Finite-frequency noise}\label{sec:noise}

We have analytically derived $S^{\pm} (\Omega)$ up to second order in the poor man's scaling solution  $J=J(\Lambda_c)$. In the scaling limit ($D\to\infty, J_0\to 0$ at fixed $T_K$) and for $\Omega\gg T_K$ they read
\begin{subequations}
\label{eq:S}
\begin{align}
\label{eq:Sp_final}
&\begin{aligned}[b]&S^+(\Omega)=\pi J_\nd^2Mh+\frac{\pi}{8}J_\nd^2\sum_{\alpha,\s=\pm}
|\Omega+\alpha V+\s  h|_2\\
&\quad+\frac{\pi}{2} J_\nd^2\sum_{\alpha=\pm}\left[M^2|\Omega+\alpha V|-\left(M^2-\frac{1}{4}\right)|\Omega+\alpha V|_1\right], 
\end{aligned}\\
\label{eq:Sm_final}
&S^-(\Omega)=\frac{3\pi}{4}J_\nd^2\Omega+\frac{\pi}{4}J_\nd^2M\sum_{\alpha,\s=\pm}\s
|\Omega+\alpha V+\s  h|_2\,,
\end{align}
\end{subequations}
where $|x|_i=(2x/\pi)\arctan(x/\Gamma_i)$ is the absolute value function smeared on the scale $\Gamma_i$. The longitudinal and transverse relaxation rates are given by $\Gamma_1=\pi(J_L^2+J_R^2)|h|/2+\pi J_\nd^2\max\{|V|,|h|\}$ and $\Gamma_2=\pi J_\nd^2|V|/2+\Gamma_1/2$ respectively~\cite{PhysRevB.80.045117}. The dot magnetization is $M=-(1+r)^2h/[2(1+r^2)|h|+4r\max\{|V|,|h|\}]$, with the renormalized magnetic field $h=(1-J)h_0$ and the asymmetry $r=x_L/x_R$. 

We stress that the RTRG method provides a consistent derivation of the relaxation rates appearing in Eqs.~\eqref{eq:S} via the smeared absolute value function. In particular, these are absent in bare 
second-oder perturbation theory which is obtained by replacing the renormalized exchange coupling $J$ with the bare one $J_0$ and taking the limit $\Gamma_i \to 0$. We note that in this limit the contribution proportional to $M^2$ in $S^+$ vanishes, which within the RTRG analysis introduces new effects discussed in the following.

\begin{figure}[tb]
	\centering
	\includegraphics[width=\linewidth,clip=true]{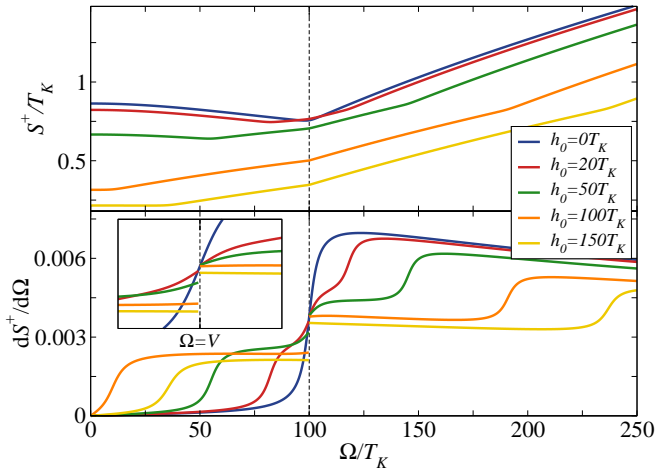}
		\caption{(Color online) Upper panel: Symmetric noise $S^+(\Omega)$ for 
		$V=100\,T_K$, $r=1$, and 
		different magnetic fields. Lower panel: 
		Derivative $\di S^+/\di\Omega$ showing a discontinuous jump at $\Omega=V$ for finite 
		magnetic fields (zoom in the inset), see Ref.~\onlinecite{note1}.}
		\label{fig:fig2}
\end{figure}
\begin{figure}[b]
	\centering
		\includegraphics[width=\linewidth,clip=true]{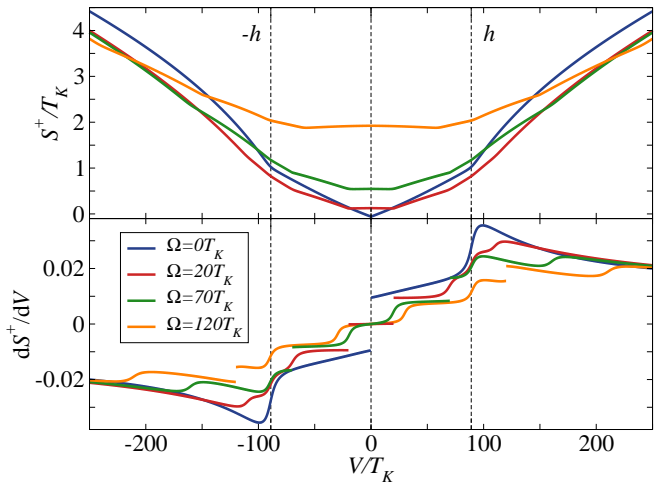}
		\caption{(Color online) Symmetric noise $S^+(\Omega)$ as a function of bias voltage $V$ for magnetic field $h_0=100\,T_K$ leading to $h=89\,T_K$, $r=1$, and different frequencies (upper panel), and derivative $\di S^+(V)/\di V$ (lower panel), see Ref.~\onlinecite{note1}.}
	\label{fig:fig3}
\end{figure}
\begin{figure}[tb]
	\centering
		\includegraphics[width=\linewidth,clip=true]{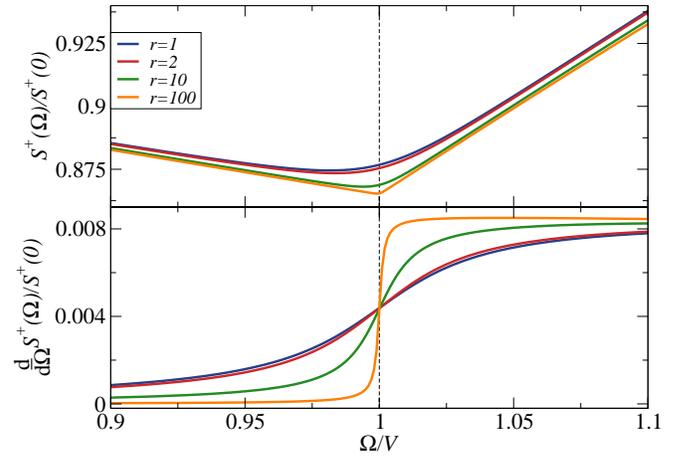}
		\caption{(Color online) Normalized symmetric noise $S^+(\Omega)/S^+(0)$ for $V=100\,T_K$, $h_0=0$, and different asymmetries $r$ (upper panel), and derivative $\dd{}{\Omega}S^+(\Omega)/S^+(0)$ (lower panel), see Ref.~\onlinecite{note1}.}
	\label{fig:fig4}
\end{figure}

\subsection{Symmetric noise}

We plot the symmetric noise and its derivative~\cite{note1} in
Fig.~\ref{fig:fig2}. We consider $r = 1$, the asymmetry effects are discussed below. 
For vanishing magnetic field Eq.~\eqref{eq:Sp_final} simplifies to 
\begin{align}
S^+(\Omega)=\frac{3\pi}{8} J_\nd^2\sum_{\alpha=\pm}|\Omega+\alpha V|_{\Gamma}\,,
\end{align}
where $|x|_\Gamma =(2x/\pi)\arctan(x/\Gamma)$ with $\Gamma =\Gamma_1=\Gamma_2=\pi J_\nd^2V$ for $h_0=0$. 
The pronounced feature at $\Omega = V$ (blue curve) leads to a characteristic resonance in
the derivative. 
For finite magnetic fields additional features at 
$\Omega = |V \pm h|$ arise, which yield enhancements
in the derivative broadened by the transverse spin relaxation rate $\Gamma_2$.
Furthermore, due to the 
terms proportional to $M^2$ in \eqref{eq:Sp_final} 
at $\Omega=\pm V$ we find a contribution to $S^+$ which is not broadened by any microscopic decay rate. As shown in Fig.~\ref{fig:fig2} 
the sharp kink at $V=\Omega$ 
yields a \emph{discontinuity} in the derivative with the jump given by
\begin{equation}
\Delta=\pi J_\nd^2M^2\,.
\label{eq:Delta}
\end{equation}
For $h < V$ the contribution proportional
to $(M^2-1/4)$ provides a superposition with a continuous
enhancement at $\Omega = V$ broadened by $\Gamma_1$. This
dependence on the magnetic field is shown in the inset of
Fig.~\ref{fig:fig2}.
We note that the singular behavior can already be observed in equilibrium (see Fig.~\ref{fig:fig3}). 
For zero voltage and finite magnetic field Eq.~\eqref{eq:Sp_final} reads
\begin{align}
S^+(\Omega)=&\frac{\pi}{4}J_\nd^2\Big[|\Omega|-2h+\sum_{\s=\pm}|\Omega+\s h|_2\Big]
\end{align}
with $\Gamma_2(V=0)=2\pi J_c^2|h|$. The absolute value at $\Omega=0$ is shifted to 
$\Omega=\pm V$ for finite voltages. 

In contrast to $S^+(\Omega)$, the antisymmetric noise $S^-(\Omega)$ contains only terms with resonances broadened by $\Gamma_2$. 
This behavior is reflected in
the ac conductance and will be discussed below.

In Fig.~\ref{fig:fig4} we consider asymmetry effects, which involve a rescaling of the exchange couplings $J_\nd^2$ by $4r/(1+r)^2$. 
The magnetization depends only weakly on $r$.
As a consequence, the reduced relaxation rates lead to a sharpening of the features close to the resonances. 

For $\Omega \gg T_K$ the irreducible contribution to $C^{\pm}(\Omega)$ (given by its first term) is dominant, while the reducible one (the second term) is subleading $\sim O(J^4)$. However, for $\Omega\ll T_K$ the reducible term contributes in order $J^2$ supplementing Eq.~\eqref{eq:Sp_final} in the limit $\Omega\to 0$ by 
\begin{equation}
-\frac{\pi^2 J_\nd^4}{2 \Gamma_1}\bigg[2 VM+\left(M^2+\frac{1}{4}\right)m(V,h)\bigg]m(V,h)\,,
\label{eq:additionalterm}
\end{equation}
with $m(V,h)=|V+h|_2-|V-h|_2$. In total, this result generalizes the nonequilibrium shot 
noise~\cite{meir} of a Kondo quantum dot to the case of finite magnetic fields.

We finally consider the noise to current~\cite{PhysRevB.80.045117} ratio $S^+(0)/I$. For $V<h$ we obtain
\begin{equation}
\frac{S^+(0)}{I}=\frac{V-2h+\sum_{\s=\pm}|V+\s h|_2}{3V-\sum_{\s=\pm} \s|V+\s h|_2}\,.
\end{equation}
In equilibrium $S^+(0)/I=1/3$ for $h\gg T_K$. 
The noise to current ratio increases with bias voltage, 
reaching the Poisson limit  $S^+(0)/I=1$ for $V\gg h$. Similar results are obtained in the Toulouse limit~\cite{PhysRevB.58.14978}.

\subsection{Fluctuation-dissipation ratio}

\begin{figure}
	\centering
		\includegraphics[width=\linewidth]{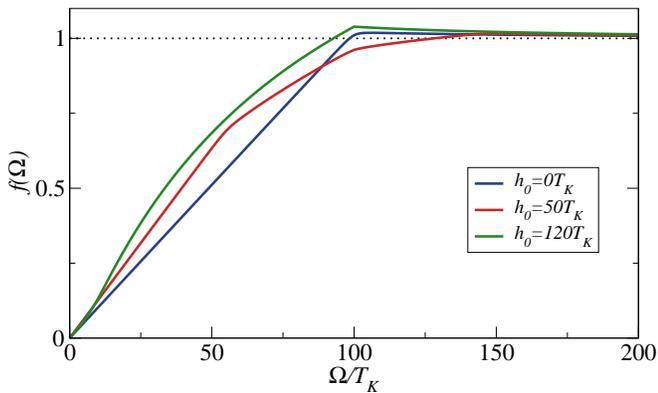}
	\caption{(Color online) Fluctuation-dissipation ratio $f(\Omega)$ for $V=100\,T_K$, $r=1$, and different magnetic fields.}
	\label{fig:fig5}
\end{figure}
In equilibrium, the symmetric and antisymmetric noise $S^{\pm}(\Omega)$ satisfy the FDT.
For vanishing magnetic field, the limit  $\Omega\rightarrow\infty$ reads
\begin{subequations}
\begin{align}
S^+(\Omega)&=\frac{3\pi}{4} J_{\rm nd}^2 |\Omega|\\
S^-(\Omega)&=\frac{3\pi}{4} J_{\rm nd}^2 \Omega\,,
\end{align}
\end{subequations}
that is Eq.~\eqref{eq:FDT} is obviously satisfied.

To investigate the violation of the FDT out of equilibrium we introduce the fluctuation-dissipation ratio
\begin{equation}
\label{fdt}
f(\Omega)=\frac{S^-(\Omega)}{S^+(\Omega)}\,,
\end{equation}
which in equilibrium is given by $f(\Omega)= {\sgn}(\Omega)$ at zero temperature. Using the results \eqref{eq:S} we obtain 
\begin{equation}
f(\Omega)=\frac{2\Omega}{\sum_{\alpha=\pm}
|\Omega+\alpha V|_{\Gamma}}\,.
\end{equation}
We find $f(\Omega> V ) = 1$, i.e.
the equilibrium result holds, whereas for small frequencies we obtain the linear behavior $f(\Omega\ll V )=\Omega/|V|_{\Gamma}$. 

At either large magnetic fields or strong asymmetries $r$, the voltage effects are suppressed, see Fig.~\ref{fig:fig5}. In particular, we note that $f(\Omega)=1$ for $\Omega  > V +h$.

\subsection{Emission noise}

From the expressions \eqref{eq:S} we determine the emission noise 
$S^e(\Omega)=S^+(\Omega)-S^-(\Omega)$ 
describing the noise induced by photon emission
\begin{equation}
\begin{aligned}[b]
S^e(\Omega)&=\pi J_\nd^2Mh-\frac{3\pi}{4}J_\nd^2\Omega\\
&+\frac{\pi}{2}J_\nd^2\!\sum_{\alpha=\pm}\!\Big[M^2|\Omega+\alpha V|\!-\!\Big(M^2-\frac{1}{4}\Big)|\Omega+\alpha V|_1\Big]\\
&+\frac{\pi}{4}J_\nd^2\sum_{\alpha,\s=\pm}\Big(\frac{1}{2}-\s M\Big)|\Omega+\alpha V+\s h|_2\,.
\end{aligned}
\label{eq:Se}
\end{equation}
It inherits the features of the symmetric noise discussed above, which can be probed in the measurements of $\di S^e(\Omega)/\di\Omega$ or $\di S^e(\Omega)/\di V$. 
\begin{figure}
	\centering
		\includegraphics[width=\linewidth,clip=true]{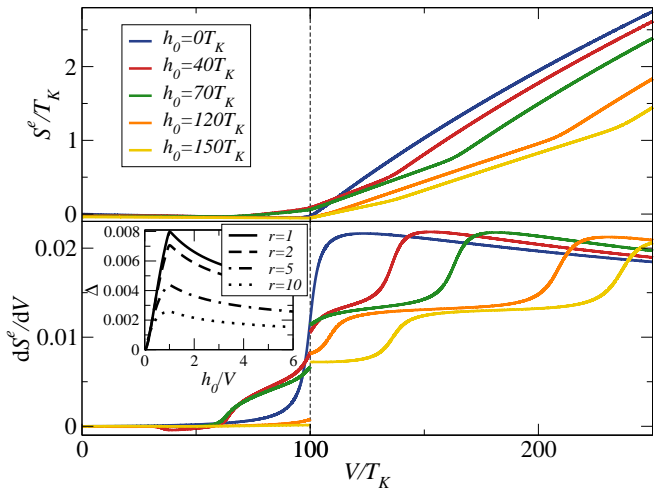}
			\caption{(Color online) Emission noise $S^e(V)$ (upper panel) and derivative $\di S^e(V)/\di V$ (lower panel)~\cite{note1} for $\Omega=100\,T_K$, $r=1$, and different magnetic fields. In the inset the dependence of the discontinuity $\Delta$ on the magnetic field is shown for $V=100\,T_K$ and different asymmetries $r$.}
	\label{fig:fig6}
\end{figure}
\begin{figure*}
	\centering
		\includegraphics[scale=1.1,clip=true]{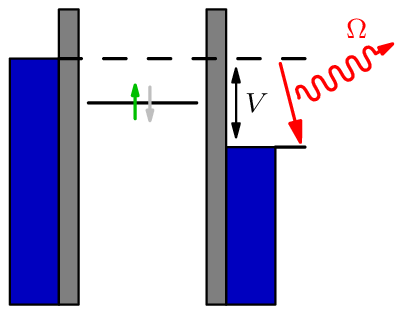}\hfil\includegraphics[scale=1.1,clip=true]{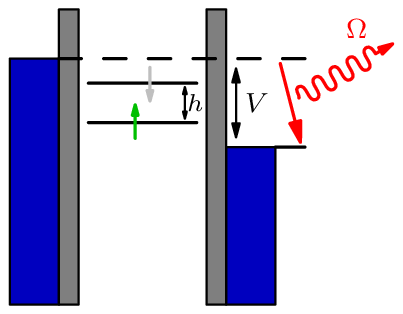}\hfil\includegraphics[scale=1.1,clip=true]{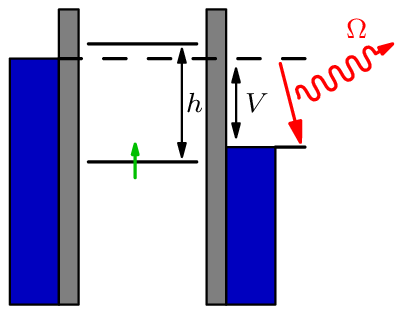}
\caption{(Color online) Visualization of the origin and behavior of the discontinuity in $\di S^e(V)/\di V$ at $V=\Omega$ for different magnetic fields.}
	\label{fig:fig7}
\end{figure*}

From Eq.~\eqref{eq:Se} the voltage and frequency dependence of $S^e(\Omega)$ appear to be very similar except for the additional features at $V=h$ due to the voltage dependence of the magnetization. As the voltage dependence is experimentally more easily accessible,
we focus on the voltage dependence shown in Fig.~\ref{fig:fig6}.
For zero magnetic field the emission noise reduces to
\begin{equation}
S^e(\Omega)=\frac{3\pi}{8} J_\nd^2\bigg(\sum_{\alpha=\pm}|\Omega+\alpha V|_{\Gamma}-2\Omega\bigg)\,,
\label{eq:Seh0}
\end{equation}
leading to a suppression for $V<\Omega$. 
At finite magnetic fields we distinguish two regimes. 
For $h<V$, the photon emission is suppressed for $V<|\Omega-h|$.
For larger magnetic fields $h>V$, the feature at $V=|\Omega-h|$ disappears for $M=-1/2$. 

The emergence of the singular behavior in the emission noise at $\Omega=\pm V$ can be attributed 
to the fact that at large fields, $h>V$, the spin on the dot is fixed to its ground state 
(see Fig.~\ref{fig:fig7}). Processes at external frequencies $\Omega=V$ probe the charge transfer between the leads, which are not broadened due to the sharpness of the Fermi edges at zero temperature. At smaller fields, $0<h<V$, the spin becomes dynamical, and virtual processes involving  longitudinal spin fluctuations give an additional, \emph{continuous}, contribution broadened by $\Gamma_1$. 
In turn, processes involving a spin flip on the dot, which appear at $\Omega=|V\pm h|$, are broadened by $\Gamma_2$. The latter behavior is also found for all resonances appearing in the current~\cite{PhysRevB.80.045117}. Thus the noise offers a way to study richer relaxation phenomena than those present in the current. We note that a discontinuity in the derivative of the noise was also found~\cite{PhysRevB.58.14978} in the strong-coupling regime of the Kondo model at the Toulouse point; we therefore expect it to be a generic feature of the finite-frequency noise in Kondo quantum dots. 

The inset of Fig.~\ref{fig:fig6} displays the jump $\Delta$ as a function of the magnetic field for different asymmetries. 
For $h<V$ the increase of $M^2$ leads to a maximum at $h=V$, while for $h>V$ the decrease of $J_\nd^2$  dominates.
The most pronounced jump is obtained for the symmetric case with $r=1$. The reduction with increasing asymmetry is inferred by the $r$ dependence of $J_\nd^2$, in addition
to the $r$ dependence of the magnetization for $h<V$. 

\section{Experimental observation}\label{sec:exp}
The emission noise describing the noise induced by photon emission~\cite{Schoelkopf} can be probed in the measurements of $\di S^e(\Omega)/\di\Omega$ or $\di S^e(\Omega)/\di V$. 
In particular, at finite magnetic field a very sharp feature is expected at $\Omega = \pm V$ which in experiments will only be broadened by finite temperatures, instrumental resolution, or charge fluctuations not captured in the Kondo model~\eqref{eq:model}.
We compare our results for the emission noise to the experimental data 
by Basset~\emph{et al.}~\cite{PhysRevLett.108.046802} at zero field and find very good agreement without adjustable parameters.
We moreover discuss the predictions for a measurement at finite magnetic fields.

\subsection{Comparison to zero field data}
We first consider vanishing magnetic field as in the recent experiments by Basset~\emph{et al.}~\cite{PhysRevLett.108.046802} on the emission noise of a carbon nanotube quantum dot in the Kondo regime. For this case Eq.~\eqref{eq:Seh0} yields
\begin{equation}\label{eq:dSLLedV_final}
\dd{S^e(\Omega)}{V}=
\frac{3}{4}J_\nd^2\left(\arctan\frac{\Omega+V}{\Gamma}-\arctan\frac{\Omega-V}{\Gamma}\right)\,,
\end{equation}
where $J_\nd^2=x_L(1-x_L)/\ln^2(\sqrt{\Omega^2+V^2}/T_K)$. 
We emphasize that Eq.~\eqref{eq:dSLLedV_final} contains only two unknown parameters, namely the Kondo temperature $T_K$ and the asymmetry $x_L$, which are extracted from the differential conductance (see below). The previous theoretical analysis of the data in Ref.~\onlinecite{PhysRevLett.108.046802} used a frequency-dependent RG analysis which required, however, the fitting of the line shape close to the resonances with phenomenological relaxation rates. Here, in contrast, the rate $\Gamma$ was derived consistently and does not contain free fit parameters.

\begin{figure}[b]
	\centering
		\includegraphics[width=\linewidth,clip=true]{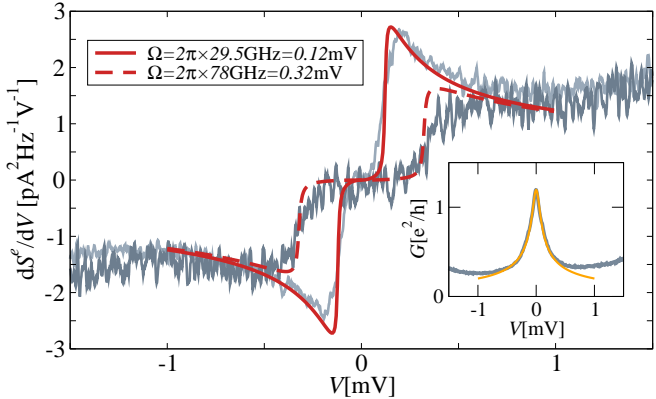}
		\caption{(Color online) Comparison of Eq.~\eqref{eq:dSLLedV_final} to the experimental 
		data of Ref.~\onlinecite{PhysRevLett.108.046802} for the derivative of the emission noise 
		$\di S^e/\di V$ at $h_0=0$. We stress that Eq.~\eqref{eq:dSLLedV_final} does not
		contain any free parameter. Inset: Fit of $G(V)$ to the theoretical result~\cite{arxiv1201.6295}.}
	\label{fig:fig8}
\end{figure}

In order to determine $T_K$ and $x_L$ we fit the measured differential conductance~\cite{PhysRevLett.108.046802} to the theory~\cite{arxiv1201.6295} (see inset of Fig.~\ref{fig:fig8}). The asymmetry is extracted~\cite{Kretinin} from $G(V=0)=1.194 \,e^2/h$ and amounts to $x_L\approx 0.82$ (or $0.18$), while the Kondo temperature is obtained from $G(V=T_K^*)=\frac23 G(V=0)$ and $T_K^*=10.57\,T_K$ \cite{factor} and equals $T_K\approx 110\,\text{mK}\approx 0.01\,\text{mV}$. 
Using these parameters we plot Eq.~\eqref{eq:dSLLedV_final} against the experimental results~\cite{PhysRevLett.108.046802} in Fig.~\ref{fig:fig8}. In the range $|V| \lesssim 1{\rm mV}$ we find excellent agreement for both frequencies; for larger voltages charge fluctuations set in, and the Kondo model \eqref{eq:model} is no longer adequate. For this reason our analysis of the features at $V=\pm\Omega$ is limited to $\Omega\lesssim 1\text{mV}$. On the other hand, it is restricted by the weak-coupling condition $\Omega\gg T_K\approx 0.01\,\text{mV}$, leaving two orders of magnitude in the window of admissible frequencies.

\subsection{Predictions for finite field}

\begin{figure}[b]
\centering
\includegraphics[width=\linewidth,clip=true]{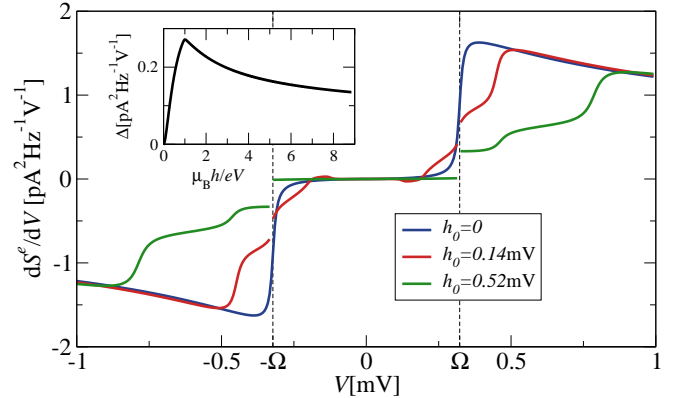}
\caption{(Color online) Voltage derivative of the emission noise for the experimental parameters of Ref.~\onlinecite{PhysRevLett.108.046802} with $\Omega/2\pi=78\,\text{GHz}$ (blue line). For comparison we show the result for small (red line) and large (green line) magnetic fields. For finite fields we observe discontinuous jumps at $V=\pm\Omega$. The relation between $h_0$ and the applied field is given by $h_0=g^*h_\text{app}/2$ with the material specific effective g-factor $g^*$. Inset: Magnetic-field dependence of the jump height.}
\label{fig:fig9}
\end{figure}

We propose to measure the emission noise of a quantum dot in the Kondo regime at finite magnetic field (see Fig.~\ref{fig:fig9}). As can be easily inferred from Eq.~\eqref{eq:Se} the energy scale $h$ has two effects on $\di S^e(\Omega)/\di V$: (i) It introduces additional features at $V=\pm|\Omega+h|$, $V=\pm|\Omega-h|$, and $V=\pm h$, which originate from the onset of additional transport processes as well as from the voltage dependence of the dot magnetization $M$. (ii) At large magnetic fields $h>\Omega$ the resonances at $V=\pm\Omega$ turn into discontinuous jumps. At smaller fields these jumps are superimposed with a contribution broadened by the longitudinal spin relaxation rate $\Gamma_1$, while all other resonances are broadened by the transverse rate $\Gamma_2$. For illustration we show $\di S^e(\Omega)/\di V$ for the parameters of Ref.~\onlinecite{PhysRevLett.108.046802} but finite magnetic fields in Fig.~\ref{fig:fig9}. In experiments the jumps at $V=\pm\Omega$ will be broadened by finite temperature $T$. This broadening is linear in $T$, in contrast to the other resonances which are broadened by $\Gamma_{2}+\mathcal{O}(T)$. For this reason the features at $V=\pm\Omega$ stay much sharper than all other ones as long as $T\ll \Gamma_{2}\sim T_K$. 

\section{AC conductance}\label{sec:ac}

Finally we discuss the nonequilibrium ac conductance. We consider a setup~\cite{Kaminski} at finite dc bias $V$ modulated by a small ac voltage $\delta V$, $V(t)=V+\delta V\,e^{-i\Omega t}$. This induces a frequency-dependent current $I(V,\delta V,\Omega)$ from which the nonequilibrium ac conductance can be extracted via $G(\Omega) = \lim_{\delta V \to 0} \frac{1}{\delta V} [I(V,\delta V,\Omega)-I(V)]$ with $I(V)$ denoting the stationary dc current. We stress that $G(\Omega)$ is the ac conductance in a nonequilibrium stationary state, i.e. in the presence of the finite dc bias $V$. 

Using Eq.~\eqref{eq:G}, the real and imaginary part of $G(\Omega)$ are determined by 
$C^- (\Omega)$. As a consequence, the singular behavior in $S^+(\Omega)$ due to the terms characterized by the absence of any decay rate is not reflected in the ac conductance.
To second order in the renormalized exchange coupling we find 
\begin{subequations}
\label{eq:G_final}
\begin{align}
\label{eq:ReG_final}
&\Re G(\Omega)=\frac{3\pi}{4}J_\nd^2
+\frac{\pi M}{4\Omega}J_\nd^2\sum_{\alpha,\s=\pm}\s\,|\Omega+\alpha V+\s h|_2\\
\label{eq:ImG_final}
&\begin{aligned}[b]
\Im\,G(\Omega)=-\frac{M}{2\Omega}J_\nd^2\sum_{\alpha,\s=\pm}\s\,
\Bigl[&\mathcal{L}_2(\Omega+\alpha V+\s h)\\
&-\mathcal{L}_2(\alpha V+\s h)\Bigr]\,,
\end{aligned}
\end{align}
\end{subequations}
where $\mathcal{L}_2(x)=x\,\ln(\Lambda_c/\sqrt{x^2+\Gamma_2^2})$ 
gives rise to logarithmic behavior at the resonances, which is a characteristic feature of Kondo systems. 
Similarly to $S^+ (\Omega)$,  for small frequencies $\Re G (\Omega)$ is supplemented by the contribution $\frac{\pi}{2} J_\nd^2m(V,h)\, (\partial M/\partial V)$ for $V>h$, where $\partial M / \partial V = - M\Gamma_1^{-1} (\partial \Gamma_1/\partial V)$, originating from the reducible part of $C^- (\Omega)$. 
We note that for the conductance this contribution is more pronounced 
due to the additional factor of $1/\Omega$ in~\eqref{eq:G}.
Thus in the limit $\Omega \to 0$ one recovers the nonequilibrium dc conductance~\cite{PhysRevB.80.045117}, while for $V \to 0$ one obtains the equilibrium ac conductance~\cite{sindel}
\begin{subequations}
\begin{align}
&\Re G_{\rm eq}(\Omega)=\frac{3\pi}{4}J_\nd^2+\frac{\pi M}{2\Omega}J_\nd^2\sum_{\s=\pm}\s\,|\Omega+\s h|_2\, , \\
&\Im G_{\rm eq}(\Omega)=-\frac{M}{2\Omega}J_\nd^2\sum_{\s=\pm}\s\,
\Bigl[\mathcal{L}_2(\Omega+\s h)-\mathcal{L}_2(\s h)\Bigr]\, ,
\end{align}
\end{subequations}
with features at $\Omega=\pm h$.
We observe that in contrast to the symmetric noise~\eqref{eq:G_final} presents no feature at $\Omega = V$ for finite magnetic fields. Moreover, all resonances are broadened by the transverse rate $\Gamma_2$, thus the finite-frequency noise contains more information on the relaxation processes than the conductance.
\begin{figure}[tb]
	\centering
		\includegraphics[width=\linewidth,clip=true]{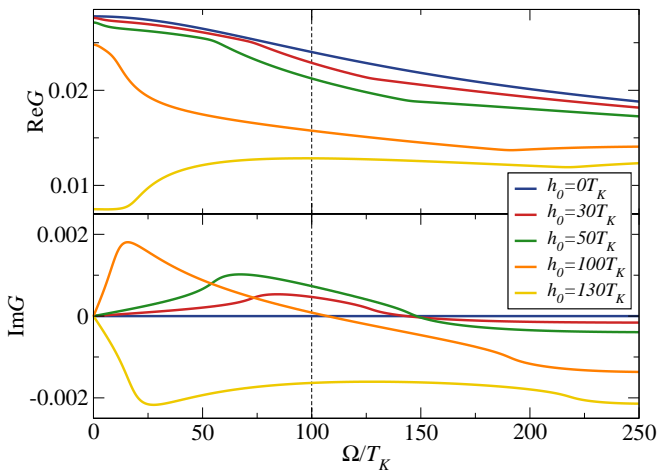}
		\caption{(Color online) Real and imaginary part of the ac
conductance $G(\Omega)$ \cite{note2} for $V = 100 \,T_K$, $r=1$, and different magnetic
fields $h_0$. We observe no feature at $\Omega = V$.}
	\label{fig:fig10}
\end{figure}

For zero magnetic field~\eqref{eq:ReG_final} simplifies to $\Re G(\Omega)=3\pi J_\nd^2/4$, the imaginary part vanishes. The frequency and voltage dependence of the real part originates in the 
$\Lambda_c$-dependence of the renormalized exchange coupling $J_{\rm nd}$.
This behavior is displayed in Fig.~\ref{fig:fig10} (blue curve), where we show results for the 
frequency dependence of $G(\Omega)$ at finite bias voltage. 
At finite magnetic field additional features appear close to the
resonances at $\Omega=|V\pm h|$ in both the real and imaginary part. 

\begin{figure}[bt]
\centering
\includegraphics[width=\linewidth,clip=true]{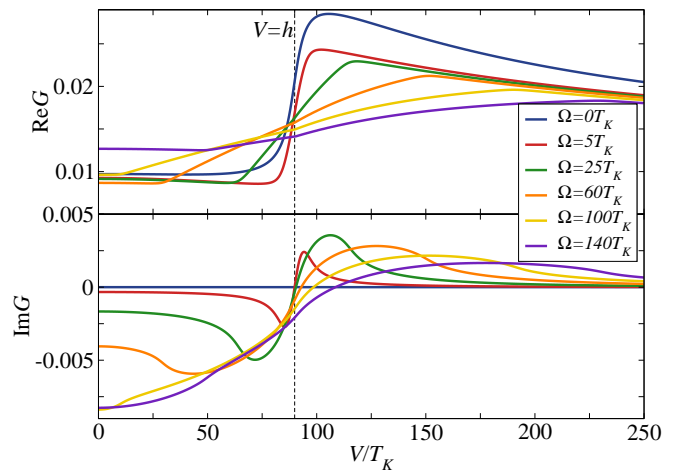}
\caption{(Color online) Nonequilibrium ac conductance $G(\Omega)$ \cite{note2} for $h_0=100\,T_K$, $r=1$, and different frequencies $\Omega$.}
\label{fig:fig11}
\end{figure}

The voltage dependence of $G(\Omega)$ at fixed frequency
is shown in Fig.~\ref{fig:fig11} 
and exhibits a qualitatively similar behavior.
The real part exhibits a characteristic enhancement at $V=h$ due to the onset of inelastic cotunneling processes. For finite frequencies this step-like enhancement is replaced by a continuous increase in the range $V=|h\pm\Omega|$ with reduced height. 
The effect of the additional contribution for $\Omega\to0$ is clearly visible.
The slight change in the slope at $V=h$ is due to the voltage dependence of the dot magnetization. 
The imaginary part, shown in the lower panel, vanishes for $\Omega = 0$. For small frequencies 
$\Omega < h$ the lineshape is approximately antisymmetric around $V = h$ except for
the offset at zero voltage.

\section{Conclusion}\label{sec:conclusion}

We have 
studied the effects of a finite magnetic field on the finite-frequency current noise and nonequilibrium ac conductance of a Kondo quantum dot. 
Using the RTRG, we present analytic solutions of the flow equations in the weak-coupling regime. 
These exhibit a novel contribution in the symmetric noise $S^+$, characterized by the absence of any decay rate as microscopic cutoff scale. 
Due to the interplay of the different energy scales both observables exhibit various resonances, close to which the lineshapes are governed by self-consistently derived decay rates. 
In particular, at finite magnetic field the symmetric noise possesses a sharp feature at $\Omega=\pm V$ 
resulting in discontinuous derivatives with respect to frequency or bias voltage.
We propose to measure the emission noise of a Kondo quantum dot at finite magnetic field, for which we have derived the full line shape including the characteristic resonances and a discontinuous jump in its derivative.

The extension of the present results to the regime of strong coupling~\cite{arxiv1201.6295} represents an interesting question to address in future investigations.

\section{Acknowledgments}

We thank W. Belzig, F. Hassler, F. Haupt, A. Komnik, V. Meden, H. Schoeller, P. Simon, J. Splettstoesser, and G. Zar\'and for valuable discussions. We are particularly grateful to R. Deblock for the insights from the experimental side and for providing us with the finite-frequency noise data. This work was supported by the Deutsche Forschungsgemeinschaft through FOR 723, FOR 912, and the Emmy-Noether Program (D.S.).


\appendix
\section{Generalization of the Kubo formula to nonequilibrium distributions}
\label{app:a}

We provide here the derivation of Eq.~\eqref{eq:G}, for a setup with a small ac voltage $\delta V$
modulating the dc bias $V$ by $V(t)=V+\delta V\,e^{-i\Omega t}$. 
We split the Hamiltonian in its time-independent part $H_0$ and the 
perturbation 
\begin{equation}
H_1(t)=h_1\delta Ve^{-i\Omega t}\,,
\end{equation}
with $h_1=\frac{1}{2}\sum_\alpha \alpha N_\alpha$. To calculate the average current induced by $H_1$ to linear order in $\delta V$, we determine the density matrix to the same order. Expanding $\rho(t)=\rho^{(0)}(t)+\rho^{(1)}(t) +\cdots$ in power series of $\delta V$, we obtain the set of von Neumann equations for contributions of every order
\begin{subequations}
\begin{align}
\label{eq:dglrho_0}
	\dd{}{t}\rho^{(0)}(t)&=-i\kom{H_0}{\rho^{(0)}(t)}\\
	\label{eq:dglrho_1}
	\dd{}{t}\rho^{(1)}(t)&=-i\kom{H_0}{\rho^{(1)}(t)}-i\kom{H_1(t)}{\rho^{(0)}(t)}\,.
\end{align}
\end{subequations}
The first equation is solved by the density matrix of the unperturbed system in Heisenberg representation 
\begin{equation}
\rho^{(0)}(t)=U_0(t)\rho_0U_0\da(t)\,, \quad U_0(t)=e^{-iH_0t}\,,
\end{equation}
where $\rho_0$ is the initial density matrix. To solve the equation for $\rho^{(1)}(t)$, we introduce a correction $U_1 (t)$ to the unperturbed time evolution operator $U_0 (t)$, which includes the linear effects of the perturbation 
\begin{align}
U_1(t)&=\nonumber\\
&\hspace{-0.75cm}-iU_0(t)\int_{0}^t\di t\p e^{-i\Omega t\p}U_0\da(t\p )h_1U_0(t\p)\,.
\end{align}
Assuming that the initial density matrix $\rho_0$ is independent of $\delta V$, we express
\begin{equation}
\rho^{(1)}(t)=U_1(t)\rho_0U_0\da(t)+U_0(t)\rho_0U_1\da(t)\,.
\label{eq:rho_1}
\end{equation}
The average current induced by the perturbation to linear order in $\delta V$ is then given by
\begin{equation}
\Ew{I_\alpha^{(1)}}(t)=\Tr\{ I_\alpha^{(0)}\rho^{(1)}(t)\}\,,
\label{eq:ewI}
\end{equation}
as $I_\alpha^{(1)}=-i\kom{H_1(t)}{N_\alpha}=0$.
Using Eq.~\eqref{eq:rho_1} and the relation $I^{(1)}(t) =G(\Omega,t)e^{-i\Omega t}\delta V$ we obtain  the ac conductance
\begin{align}
\hspace{-.1cm}G(\Omega,t)&=-ie^{i\Omega t}\int_{0}^t\di t\p e^{-i\Omega t\p}\Tr\{\,\kom{I_\alpha^{(0)}(t)}{h_1(t\p)}\rho_0\}\nonumber\\
&\hspace{-.7cm}=\frac{1}{\Omega}\Tr\{\,\kom{I_\alpha^{(0)}(t)}{h_1(t)}\rho_0\}-\frac{e^{i\Omega t}}{\Omega}\Tr\{\,\kom{I_\alpha^{(0)}(t)}{h_1}\rho_0\}\nonumber\\
&\hspace{-.35cm}-\frac{1}{\Omega}\int_{0}^t\di t\p e^{i\Omega(t-t\p)}\Tr\{\,\kom{I_\alpha^{(0)}(t)}{\dd{}{t\p}h_1(t\p)}\rho_0\}\,,
\end{align}
where we performed an integration by parts. The second term vanishes as the system is initially decoupled and $\kom{h_1}{\rho_0}=0$. By virtue of 
\begin{equation}
\dd{}{t\p}h_1(t\p)=i\kom{H_0}{h_1(t\p)}=-I^{(0)}(t\p) \,,
\end{equation}
 $G(\Omega,t)$ can be expressed as a commutator of the current operators at different times 
\begin{equation}
G(\Omega,t) =\frac{1}{\Omega} \int_{0}^{t} \!\!\di t\p (e^{i\Omega(t-t\p)}-1)\Tr\{\,\kom{I_\alpha^{(0)}(t)}{I^{(0)}(t\p)}\rho_0\}\,.
\end{equation}
In the stationary limit $t\rightarrow\infty$ the ac conductance is determined by the antisymmetric current-current correlation function $C^-(\Omega)$
\begin{align}
	G(\Omega)&=\frac{1}{\Omega}\int_{-\infty}^0\di t (e^{-i\Omega t}-1)\Ew{\kom{I^{(0)}(0)}{I^{(0)}(t)}}\nonumber\\
&=\frac{1}{\Omega}[C^-(\Omega)-C^-(0)]\,,
\end{align}
where $G(\Omega)$  is the ac
conductance in a non-equilibrium stationary state, i.e. in
presence of the finite dc bias $V$.

\section{RTRG analysis of the current-current correlation function}
\label{sec:rtrg}

In this section we set up generic RG equations for the calculation of the current-current correlation function for a model of a quantum dot coupled to electronic leads with spin and/or orbital fluctuations. We extend the calculations for dynamical correlation functions of Ref.~\onlinecite{PhysRevB.80.075120} to current-current correlations. For completness we will report the basic ideas of the RTRG, with the definitions and notations of Refs.~\onlinecite{EPJST,PhysRevB.80.045117} in the first subsection. Afterwards we will present the calculation of the irreducible part of the current-current correlation function $C^\pm(\Omega)$~\eqref{eq:C} in detail. The reducible part of Eq.~\eqref{eq:C} will be discussed in the subsection~\ref{ssec:LFL}, as it contributes only in the zero frequency limit.

\subsection{Basic definitions}

We consider the reduced density matrix $\rho_\D(t)$, which is obtained from the full density matrix $\rho(t)$ by tracing out the reservoir degrees of freedom $\rho_\D(t)=\Tr_\res\rho(t)$. The full density matrix is given by the solution of the von Neumann equation 
\begin{equation}
	\rho(t)=e^{-iHt}\rho(0)e^{iHt}=e^{-iLt}\rho(0)\,,
	\end{equation}
where $L=\kom{H}{\,.\,}$ is the Liouvillian acting on the operators in Hilbert space. Similarly to the Hamiltonian $H=H_\D+H_\res+V$, the Liouvillian contains corresponding contributions for the dot, the reservoirs, and the coupling of the dot to the reservoirs. For an initially decoupled system $\rho(0)=\rho_\D(0)\rho_L\rho_R$ with an arbitrary dot density matrix $\rho_\D(0)$, and the left and right reservoir described by grand-canonical distribution functions, the dynamics of the reduced dot density matrix $\rho_\D(t)$ can be obtained from the quantum kinetic equation 
\begin{equation}
\dot{\rho}_\D(t)=-iL_\D\rho_\D(t)-i\int_{0}^t\di t\p\Sigma(t-t\p)\rho_\D(t\p)\,.
\end{equation}
Here the first term describes the dynamics of the isolated dot and the kernel $\Sigma(t-t\p)$ contains all information about the dissipation due to the coupling to the reservoirs. In Laplace space this equation is solved to
 \begin{equation}
\rho_\D(z)=\int_{0}^\infty\di te^{izt}\rho_\D(t)=\frac{i}{z-L_\D^\eff(z)}\rho_\D(0)\,,
 \end{equation}
where $L_\D^\eff(z)=L_\D+\Sigma(z)$ is the effective dot Liouvillian consisting of the bare dot Liouvillian $L_\D$ and the dissipative kernel $\Sigma(z)$ encoding the relaxation and decoherence processes of the dot. The stationary state is obtained by 
\begin{equation}
\rho_\D^\st=\lim_{z\rightarrow i0+}\frac{z}{z-L_\D^\eff(z)}\rho_\D(0)
\end{equation}
in Laplace space.
The kernel $\Sigma(z)$ is determined by a diagrammatic expansion in the interaction between the dot and the reservoirs. 

In the following we report the definitions of Refs.~\onlinecite{EPJST,PhysRevB.80.045117}. 
The interaction vertex $G_{11\p}^{pp\p}$ is defined via the interaction part of the Liouvillian
\begin{equation}
L_V=\frac{1}{2}p\p G^{pp\p}_{11\p}:J^p_1J^{p\p}_{1\p}:\,,
\end{equation}
where we implicitly sum over the indices $1=\eta\alpha\omega$ and the Keldysh indices $p,p\p=\pm$. $J^p_1$ is a quantum field superoperator in Liouville space of the reservoirs
\begin{equation}
J^p_1=\begin{cases}a_1C&\text{for }p=+\\
Ca_1&\text{for }p=-
\end{cases}\,,
\end{equation}
where $C$ is an arbitrary reservoir operator; $a_1$ is a creation/annihilation operator for $\eta=+/-$; and $G^{pp\p}_{11\p}$ is a superoperator acting on the states of a quantum dot, defined by
\begin{equation}
G^{pp\p}_{11\p}=\begin{cases}
g_{11\p}C&\text{for }p=+\\
-Cg_{11\p}&\text{for }p=-
\end{cases}\,.
\end{equation}
The contractions 
are represented by
\begin{equation}
\gamma_{11\p}^{pp\p}=p\p\Ew{J_1^pJ_{1\p}^{p\p}}_{\rho_\res}=\delta_{1\1\p}\rho(\omega)p\p f_\alpha(\alpha p\p\omega)\,,
\end{equation}
with $\rho(\omega)$ being the density of states and $f_\alpha(x)$ being the Fermi function of reservoir $\alpha$. The free propagation of the system 
between two interaction vertices is described by resolvents of the form 
\begin{equation}
\Pi(z)=\frac{1}{z-L_\D^\eff(z)}\,.
\end{equation}
 An exact derivation of the diagrammatic rules can be found in Ref.~\onlinecite{EPJST}. 

The stationary current 
\begin{equation}
\Ew{I}_\st=-i\lim_{z\rightarrow0^+}\Tr_\D\Sigma_I(z)\rho_\D^\st(z)
\end{equation}
and the current-current correlation function \eqref{eq:def_C} 
can be expressed in terms of corresponding current kernels after integrating out the reservoir degrees of freedom~\cite{PhysRevB.80.075120}. Here $\Sigma_I(\Omega)$ is the current kernel corresponding to the current operator $I_{11\p}^{pp\p}$, $\Sigma_I^\pm(\Omega,\xi)$ to the current vertex $(I^{\pm})_{11\p}^{pp\p}$, and $\Sigma_\II^\pm(\Omega,\xi)$ to the vertex $(\II^\pm)_{11\p}^{pp\p}$ with both current operators. The current operators are defined in the same way as the Liouvillian, via the commutator and anticommutator $L_I=\frac{i}{2}\akom{I}{\cdot}$ and $L_{I^\pm}=i\kompm{I}{\cdot}$, related to the interaction vertex $G^{pp\p}_{11\p}$ by
\begin{subequations}
\label{eq:def_I}
\begin{align}
\label{eq:I}
&I^{pp\p}_{11\p}=c_{11\p}^L\delta_{pp\p}pG^{pp}_{11\p} \,, \\
\label{eq:Ip}
&(I^+)^{pp\p}_{11\p}=2c_{11\p}^L\delta_{pp\p}pG^{pp}_{11\p} \,, \\
\label{eq:Im}
&(I^-)^{pp\p}_{11\p}=2c_{11\p}^L\delta_{pp\p}G^{pp}_{11\p}\,,
\end{align}
\end{subequations}
where $c_{11\p}^L=-\frac{1}{2}(\eta\delta_{\alpha L}+\eta\p\delta_{\alpha\p L})$ accounts for the antisymmetry in the lead indices. In the diagrammatic expansion it is important to distinguish between theses current vertices, since the first one \eqref{eq:I} has to be at the leftmost position of a diagram, while the other two can be at arbitrary positions. Furthermore, the current vertex $I^\pm$ acts as a separator between the two frequencies $\Omega$ and $\xi$ occurring in Eq.~\eqref{eq:C}: In front of $I^\pm$ the variable $\Omega$ of the Fourier transform of the current-current correlation function occurs in the respective resolvents, and after $I^\pm$ it is replaced by the Laplace variable $\xi$, which is later sent to zero for the stationary state.

Eq.~\eqref{eq:C} consists of two different terms $C^\pm=C^\pm_{\text{red}}+C^\pm_{\text{irr}}$. The first one is reducible with respect to the current vertices, and it is composed of two individual current kernels each containing only one current vertex. The second one is irreducible and given by the current-current kernel $\Sigma_{\II}^\pm$ containing all irreducible diagrams incorporating both current vertices. For the calculation of the current noise up to second order in the interaction between the quantum dot and the leads, we introduce a dimensionless coupling constant $J$, which fulfills $G_{11\p}^{pp\p}\propto J$. Since all kernels contain at least two vertices, all are of order $J^2$. Thus the irreducible term proportional to $\Sigma_{\II}^\pm$ always contributes, while in general the reducible one is of higher order. Only in the limit $\Omega\rightarrow0$ it might be possible that the resolvent $[\Omega-L_\D^\eff(\Omega)]^{-1}$ becomes of the order $J^{-2}$ and thus reduces the order of this term. For the Kondo model, which will be discussed in Appendix~\ref{sssec:isoKondo}, this is indeed the case. However, since the reducible term only contributes in the low frequency limit, we will focus on the determination of the current-current kernel $\Sigma_{\II}^\pm$ in the next section, while the additional $\Omega\rightarrow0$ contribution will be discussed in Appendix~\ref{ssec:LFL}.

\subsection{Finite-frequency current-current correlation function}

In this subsection we first set up the RG equations for the current-current kernel $\Sigma_\II^\pm(\Omega,\xi)$ and the required vertices for an arbitrary model with spin/orbital fluctuations. These equations are solved explicitly for the isotropic Kondo model. We finally derive the ac conductance from the finite-frequency current-current correlation function.
\subsubsection{Generic RG equations of the current-current kernel}

The RG procedure presents two steps. In the first step the symmetric part of the reservoir contractions $\gamma^{pp\p}_{11\p}$ is integrated out in a discrete step. In the second one a cutoff $\Lambda$ is introduced in the contractions via the Fermi function.

\begin{figure}%
\centering
\includegraphics[scale=1.3]{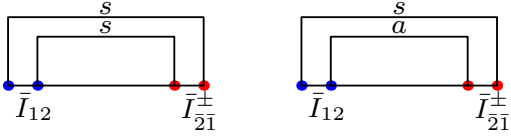}
\caption{(Color online) Diagrams for the integration of the symmetric part for the current-current kernel $\Sigma_{\II}^\pm$. The two adjacent dots symbolize the two reservoir field operators belonging to one vertex. $s(a)$ denotes the symmetric (antisymmetric) contraction $\gamma^s(\gamma^a).$}%
\label{fig:discretestep}%
\end{figure}

\emph{Discrete Step.}
For the discrete step we split the reservoir contraction into a symmetric and an antisymmetric part 
\begin{equation}
\gamma^{pp\p}_{11\p}=\delta_{1\1\p}p\p \gamma^s_1+\delta_{1\1\p}\gamma^a_1\,,
\label{eq:gamma}
\end{equation}
where $\gamma^s_1=\frac{1}{2}\rho(\bo)\ \text{and}\ \gamma^a_1=\rho(\bo)\left[f_\alpha(\bo)-\frac{1}{2}\right]$ with $\bo=\eta\omega$. In Fig.~\ref{fig:discretestep} we show the lowest order diagrams for the discrete RG step for the kernel $\Sigma_{\IAB}^\pm$. Using the diagrammatic rules \cite{EPJST} and the decomposition \eqref{eq:gamma} yields 
\begin{widetext}
\begin{subequations}
\begin{align}
	\bI_{11\p}^{\pm a (2)}&=\int\limits_{-\infty}^\infty \di\bo_2p\p\gamma^s_2(I^\pm)^{pp}_{12}\frac{1}{E_{12}+\bo_{1}+\bo_2-L_\D}G^{p\p p\p}_{\21\p}+\int\limits_{-\infty}^\infty \di\bo_2p\p\gamma^s_2G^{pp}_{12}\frac{1}{E_{12}+\bo_{1}+\bo_2-L_\D}(I^\pm)^{p\p p\p}_{\21\p}-(1\leftrightarrow1\p) \,,\\
\bIAB_{11\p}^{\pm a}&=p\p\gamma^s_2I^{pp}_{12}\frac{1}{E_{12}+\bo_{12}-L_\D}(I^\pm)^{p\p p\p}_{\21\p}-(1\leftrightarrow1\p) \,, \\
\Sigma_{\IAB}^{\pm a}&=\int\limits_{-\infty}^\infty\di \bo_1\int\limits_{-\infty}^\infty\di \bo_{1}\p\left(\frac{1}{2}\gamma^s_{1\p}+p\p\gamma_{1\p}^a\right)\gamma_1^s I^{pp}_{11\p}\frac{1}{E_{11\p}+\bo_{1}+\bo_{1}\p-L_S^{(0)}}(I^\pm)^{p\p p\p}_{\bar{1\p}\1}\,.
\end{align}
\end{subequations}
Performing the frequency integrations\cite{EPJST} and neglecting terms of order $1/D$ we obtain
\begin{subequations}
\begin{align}
\label{eq:ic_IB}
\bI^{\pm a}_{11\p}&=\bI^\pm_{11\p}\!-\!i\frac{\pi}{2}\!\left(\bI^\pm_{12}\wt{G}_{\21\p}-\bI^\pm_{1\p2}\wt{G}_{\21}+\bG_{12}\wt{I}^\pm_{\21\p}-\bG_{1\p2}\wt{I}^\pm_{\21}\right)\!,\\
\label{eq:ic_IAB}
\bIAB^{\pm a}_{11\p}&=-i\frac{\pi}{2}(\bI_{12}\wt{I}^{\pm}_{\21\p}-\bI_{1\p2}\wt{I}^{\pm}_{\21}) \,, \\
\label{eq:ic_Sigma}
\Sigma_{\II}^{\pm a}&=-i\frac{\pi^2}{16}D\bI_{11\p}\bI^{B\pm}_{\bar{1\p}\1}-i\frac{\pi}{4}\bI_{11\p}\left(E_{11\p}-L_S^{(0)}\right)\wt{I}^{\pm}_{\bar{1\p}\1}+\frac{\pi^2}{32}\bI_{11\p}\left(E_{11\p}-L_S^{(0)}\right)\bI^{\pm}_{\bar{1\p}\1}-\frac{\pi}{4}D\bI_{11\p}\wt{I}^{\pm}_{\bar{1\p}\1}\,.
\end{align}
\end{subequations}
where we used the averaged vertices $\bG_{11\p}=\sum_pG^{pp}_{11\p}$ and $\wt{G}_{11\p}=\sum_ppG^{pp}_{11\p}$ (analog for the current vertices). These results represent the initial condition for the RG equations set up in the following.

\emph{Continuous Step.}
In the continuous RG step the kernel $\Sigma_{\II}^\pm(\Omega,\omega,\xi,\xi\p)$ and the vertices $\bI_{11\p}^\pm(\Omega,\omega,\xi,\xi\p,\omega_1,\omega_{1\p})$ and $\bII_{11\p}^\pm(\Omega,\omega,\xi,\xi\p,\omega_1,\omega_{1\p})$ acquire an additional dependence on the Laplace variables $\Omega+i\omega$ and $\xi+i\xi\p$, as well as on the Matsubara frequencies $\omega_1$ and $\omega_{1\p}$. In Fig.~\ref{fig:diagramSigma} the diagrams for the current kernel $\Sigma_{\II}^\pm$ are shown. According to the diagrammatic rules developed in Ref.~\onlinecite{EPJST} we determine the RG equations for the kernel and the vertices
\begin{figure*}[tb]
	\centering
		\includegraphics[scale=1.3]{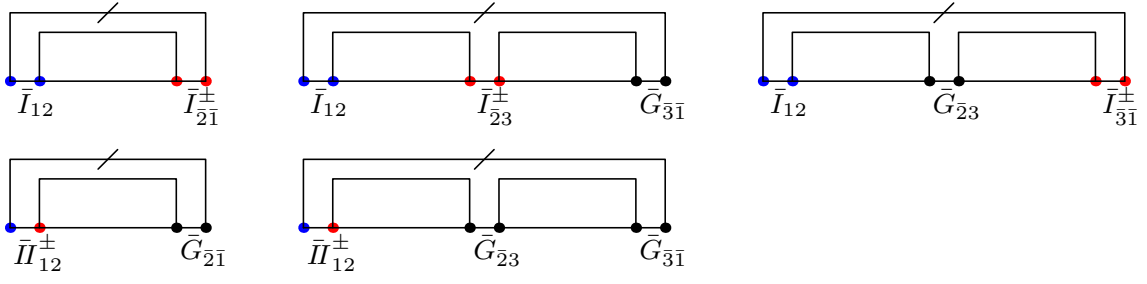}
	\caption{(Color online) RG diagrams for the renormalization of the current-current kernel $\Sigma_{\II}^\pm$.}
	\label{fig:diagramSigma}
\end{figure*}

\begin{equation}
\begin{aligned}
-\frac{\di \bI^\pm_{11\p}(\Omega,\omega,\xi,\xi\p;\omega_1,\omega_{1\p})}{\di\Lambda}
=&-i\bI^\pm_{12}(\Omega,\omega,\xi,\xi\p;\omega_1,\Lambda)\Pi(\xi_{12},\xi\p+\omega_1+\Lambda)\bar{G}_{\bar{2}1\p}(\xi_{12},\xi\p+\omega_1+\Lambda,-\Lambda,\omega_{1\p})\\
&-i\bar{G}_{12}(\Omega,\omega;\omega_1,\Lambda)\Pi(\Omega_{12},\omega+\omega_1+\Lambda)\bI^\pm_{\bar{2}1\p}(\Omega_{12},\omega+\omega_1+\Lambda,\xi,\xi\p;-\Lambda,\omega_{1\p})\\
&+\bI^\pm_{12}(\Omega,\omega,\xi,\xi\p;\omega_1,\Lambda)\Pi(\xi_{12},\xi\p+\omega_1+\Lambda)\bar{G}_{1\p3}(\xi_{12},\xi\p+\omega_1+\Lambda,\omega_{1\p},\omega_3)\\
&\times\Pi(\xi_{11\p23},\xi\p+\omega_1+\omega_{1\p}+\Lambda+\omega_3)\bar{G}_{\bar{3}\bar{2}}(\xi_{11\p23},\xi\p+\omega_1+\omega_{1\p}+\Lambda+\omega_3,-\omega_3,-\Lambda)\\
&+\bar{G}_{12}(\Omega,\omega,\omega_1,\Lambda)\Pi(_{12},\omega+\omega_1+\Lambda)\bI^\pm_{1\p3}(E_{12},\omega+\omega_1+\Lambda,\omega_{1\p},\omega_3)\\
&\times\Pi(E_{11\p23},\omega+\omega_1+\omega_{1\p}+\Lambda+\omega_3)\bar{G}_{\bar{3}\bar{2}}(E_{11\p23},\omega+\omega_1+\omega_{1\p}+\Lambda+\omega_3,-\omega_3,-\Lambda)\\
&+\bar{G}_{12}(\Omega,\omega,\omega_1,\Lambda)\Pi(\Omega_{12},\omega+\omega_1+\Lambda)\bar{G}_{1\p3}(\Omega_{12},\omega+\omega_1+\Lambda,\omega_{1\p},\omega_3)\\
&\times\Pi(\Omega_{11\p23},\omega+\omega_1+\omega_{1\p}+\Lambda+\omega_3)\bI^\pm_{\bar{3}\bar{2}}(\Omega_{11\p23},\omega+\omega_1+\omega_{1\p}+\Lambda+\omega_3,\xi,\xi\p;-\omega_3,-\Lambda)\\
&-\bI^\pm_{23}(\Omega,\omega,\xi,\xi\p;\Lambda,\omega_3)\Pi(\xi_{23},\xi\p+\Lambda+\omega_3)\bar{G}_{\bar{3}1}(\xi_{23},\xi\p+\Lambda+\omega_3,-\omega_3,\omega_1)\\
&\times\Pi(\xi_{12},\xi\p+\omega_1+\Lambda)\bar{G}_{1\p\bar{2}}(\xi_{12},\xi\p+\omega_1+\Lambda,\omega_{1\p},-\Lambda)\\
&-\bar{G}_{23}(\Omega,\omega,\Lambda,\omega_3)\Pi(\Omega_{23},\omega+\Lambda+\omega_3)\bI^\pm_{\bar{3}1}(\Omega_{23},\omega+\Lambda+\omega_3,\xi,\xi\p;-\omega_3,\omega_1)\\
&\times\Pi(\xi_{12},\xi\p+\omega_1+\Lambda)\bar{G}_{1\p\bar{2}}(\xi_{12},\xi\p+\omega_1+\Lambda,\omega_{1\p},-\Lambda)\\
&-\bar{G}_{23}(\Omega,\omega,\Lambda,\omega_3)\Pi(\Omega_{23},\omega+\Lambda+\omega_3)\bar{G}_{\bar{3}1}(\Omega_{23},\omega+\Lambda+\omega_3,-\omega_3,\omega_1)\\
&\times\Pi(\Omega_{12},\omega+\omega_1+\Lambda)\bI^\pm_{1\p\bar{2}}(\Omega_{12},\omega+\omega_1+\Lambda,\xi,\xi\p;\omega_{1\p},-\Lambda)\\
&+(1\leftrightarrow 1\p)\\
&-\bI^\pm_{23}(\Omega,\omega,\xi,\xi\p;\Lambda,\omega_3)\Pi(\xi_{23},\xi\p+\Lambda+\omega_3)\bar{G}_{11\p}(\xi_{23},\xi\p+\Lambda+\omega_3;\omega_1,\omega_{1\p})\\
&\times\Pi(\xi_{11\p32},\xi\p+\omega_1+\omega_{1\p}+\omega_3+\Lambda)\bar{G}_{\bar{3}\bar{2}}(\xi_{11\p23},\xi\p+\omega_1+\omega_{1\p}+\omega_3+\Lambda;-\omega_3,-\Lambda)\\
&-\bar{G}_{23}(\Omega,\omega,\Lambda,\omega_3)\Pi(\Omega_{23},\omega+\Lambda+\omega_3)\bI^\pm_{11\p}(\Omega_{23},\omega+\Lambda+\omega_3,\xi,\xi\p;\omega_1,\omega_{1\p})\\
&\times\Pi(\xi_{11\p32},\xi\p+\omega_1+\omega_{1\p}+\omega_3+\Lambda)\bar{G}_{\bar{3}\bar{2}}(\xi_{11\p23},\xi\p+\omega_1+\omega_{1\p}+\omega_3+\Lambda;-\omega_3,-\Lambda)\\
&-\bar{G}_{23}(\Omega,\omega,\Lambda,\omega_3)\Pi(\Omega_{23},\omega+\Lambda+\omega_3)\bar{G}_{11\p}(\Omega_{23},\omega+\Lambda+\omega_3,\omega_1,\omega_{1\p})\\
&\times\Pi(\Omega_{11\p32},\omega+\omega_1+\omega_{1\p}+\omega_3+\Lambda)\bI^\pm_{\bar{3}\bar{2}}(\Omega_{11\p23},\omega+\omega_1+\omega_{1\p}+\omega_3+\Lambda,\xi,\xi\p;-\omega_3,-\Lambda) \,, \\
\end{aligned}
\label{eq:RG_Ipm}
\end{equation}
\begin{equation}
\begin{aligned}
-\frac{\di \bII^\pm_{11\p}(\Omega,\omega,\xi,\xi\p;\omega_1,\omega_{1\p})}{\di\Lambda}
=&-i\bI_{12}(\Omega,\omega,\omega_1,\Lambda)\Pi(\Omega_{12},\omega+\omega_1+\Lambda)\bI^\pm_{\bar{2}1\p}(\Omega_{12},\omega+\omega_1+\Lambda,\xi,\xi\p;-\Lambda,\omega_{1\p})\\
&-i\bII^\pm_{12}(\Omega,\omega,\xi,\xi\p;\omega_1,\Lambda)\Pi(\xi_{12},\xi\p+\omega_1+\Lambda)\bar{G}_{\bar{2}1\p}(\xi_{12},\xi\p+\omega_1+\Lambda,-\Lambda,\omega_{1\p})\\
&+\bI_{12}(\Omega,\omega,\omega_1,\Lambda)\Pi(\Omega_{12},\omega+\omega_1+\Lambda)\bI^\pm_{1\p3}(\Omega_{12},\omega+\omega_1+\Lambda,\xi,\xi\p;\omega_{1\p},\omega_3)\\
&\times\Pi(\xi_{11\p23},\xi\p+\omega_1+\omega_{1\p}+\Lambda+\omega_3)\bar{G}_{\bar{3}\bar{2}}(\xi_{11\p23},\xi\p+\omega_1+\omega_{1\p}+\Lambda+\omega_3,-\omega_3,-\Lambda)\\
&+\bI_{12}(\Omega,\omega,\omega_1,\Lambda)\Pi(\Omega_{12},\omega+\omega_1+\Lambda)\bar{G}_{1\p3}(\Omega_{12},\omega+\omega_1+\Lambda,\omega_{1\p},\omega_3)\\
&\times\Pi(\Omega_{11\p23},\omega+\omega_1+\omega_{1\p}+\Lambda+\omega_3)\bI^\pm_{\bar{3}\bar{2}}(\Omega_{11\p23},\omega+\omega_1+\omega_{1\p}+\Lambda+\omega_3,\xi,\xi\p;-\omega_3,-\Lambda)\\
&+\bII^\pm_{12}(\Omega,\omega,\xi,\xi\p;\omega_1,\Lambda)\Pi(\xi_{12},\xi\p+\omega_1+\Lambda)\bar{G}_{1\p3}(\xi_{12},\xi\p+\omega_1+\Lambda,\omega_{1\p},\omega_3)\\
&\times\Pi(\xi_{11\p23},\xi\p+\omega_1+\omega_{1\p}+\Lambda+\omega_3)\bar{G}_{\bar{3}\bar{2}}(\xi_{11\p23},\xi\p+\omega_1+\omega_{1\p}+\Lambda+\omega_3,-\omega_3,-\Lambda)\\
&-\bI_{23}(\Omega,\omega,\Lambda,\omega_3)\Pi(\Omega_{23},\omega+\Lambda+\omega_3)\bI^\pm_{\bar{3}1}(\Omega_{23},\omega+\Lambda+\omega_3,\xi,\xi\p;-\omega_3,\omega_1)\\
&\times\Pi(\xi_{12},\xi\p+\omega_1+\Lambda)\bar{G}_{1\p\bar{2}}(\xi_{12},\xi\p+\omega_1+\Lambda,\omega_{1\p},-\Lambda)\\
&-\bI_{23}(\Omega,\omega,\Lambda,\omega_3)\Pi(\Omega_{23},\omega+\Lambda+\omega_3)\bar{G}_{\bar{3}1}(\Omega_{23},\omega+\Lambda+\omega_3,-\omega_3,\omega_1)\\
&\times\Pi(\Omega_{12},\omega+\omega_1+\Lambda)\bI^\pm_{1\p\bar{2}}(\Omega_{12},\omega+\omega_1+\Lambda,\xi,\xi\p;\omega_{1\p},-\Lambda)\\
&-\bII^\pm_{23}(\Omega,\omega,\xi,\xi\p;\Lambda,\omega_3)\Pi(\xi_{23},\xi\p+\Lambda+\omega_3)\bar{G}_{\bar{3}1}(\xi_{23},\xi\p+\Lambda+\omega_3,-\omega_3,\omega_1)\\
&\times\Pi(\xi_{12},\xi\p+\omega_1+\Lambda)\bar{G}_{1\p\bar{2}}(\xi_{12},\xi\p+\omega_1+\Lambda,\omega_{1\p},-\Lambda)\\
&+(1\leftrightarrow 1\p)\\
&-\bI_{23}(\Omega,\omega,\Lambda,\omega_3)\Pi(\Omega_{23},\omega+\Lambda+\omega_3)\bI^\pm_{11\p}(\Omega_{23},\omega+\Lambda+\omega_3,\xi,\xi\p;\omega_1,\omega_{1\p})\\
&\times\Pi(\xi_{11\p32},\xi\p+\omega_1+\omega_{1\p}+\omega_3+\Lambda)\bar{G}_{\bar{3}\bar{2}}(\xi_{11\p23},\xi\p+\omega_1+\omega_{1\p}+\omega_3+\Lambda,-\omega_3,-\Lambda)\\
&-\bI_{23}(\Omega,\omega,\Lambda,\omega_3)\Pi(\Omega_{23},\omega+\Lambda+\omega_3)\bar{G}_{11\p}(\Omega_{23},\omega+\Lambda+\omega_3,\omega_1,\omega_{1\p})\\
&\times\Pi(\Omega_{11\p32},\omega+\omega_1+\omega_{1\p}+\omega_3+\Lambda)\bI^\pm_{\bar{3}\bar{2}}(\Omega_{11\p23},\omega+\omega_1+\omega_{1\p}+\omega_3+\Lambda,\xi,\xi\p;-\omega_3,-\Lambda)\\
&-\bII^\pm_{23}(\Omega,\omega,\xi,\xi\p;\Lambda,\omega_3)\Pi(\xi_{23},\xi\p+\Lambda+\omega_3)\bar{G}_{11\p}(\xi_{23},\xi\p+\Lambda+\omega_3,\omega_1,\omega_{1\p})\\
&\times\Pi(\xi_{11\p32},\xi\p+\omega_1+\omega_{1\p}+\omega_3+\Lambda)\bar{G}_{\bar{3}\bar{2}}(\xi_{11\p23},\xi\p+\omega_1+\omega_{1\p}+\omega_3+\Lambda,-\omega_3,-\Lambda)\,,
\end{aligned}
\label{eq:RG_IIpm}
\end{equation}
\begin{equation}
\begin{aligned}[b]
\frac{\di \Sigma_{\II}^\pm(\Omega,\omega,\xi,\xi\p)}{\di\Lambda}
=&+\bI_{12}(\Omega,\omega;\Lambda,\omega_2)\Pi(\Omega_{12},\omega+\omega_2+\Lambda)\bI^\pm_{\bar{2}\bar{1}}(\Omega_{12},\omega+\Lambda+\omega_2,\xi,\xi\p;-\omega_2,-\Lambda)\\
&+\bII^\pm(\Omega,\omega,\xi,\xi\p;\Lambda,\omega_2)\Pi(\xi_{12},\xi\p+\omega_2+\Lambda)\bar{G}_{\bar{2}\bar{1}}(\xi_{12},\xi\p+\Lambda+\omega_2,-\omega_2,-\Lambda)\\
&-i\bI_{12}(\Omega,\omega,\Lambda,\omega_2)\Pi(\Omega_{12},\omega+\omega_2+\Lambda)\bI^\pm_{\bar{2}3}(\Omega_{12},\omega+\omega_2+\Lambda,\xi,\xi\p;-\omega_2,\omega_3)\\
&\times\Pi(\xi_{13},\xi\p+\omega_3+\Lambda)\bar{G}_{\bar{3}\bar{1}}(\xi_{13},\xi\p+\omega_3+\Lambda,-\omega_3,-\Lambda)\\
&-i\bI_{12}(\Omega,\omega,\Lambda,\omega_2)\Pi(\Omega_{12},\omega+\omega_2+\Lambda)\bar{G}_{\bar{2}3}(\Omega_{12},\omega+\omega_2+\Lambda,-\omega_2,\omega_3)\\
&\times\Pi(\Omega_{13},\omega+\omega_3+\Lambda)\bI^\pm_{\bar{3}\bar{1}}(\Omega_{13},\omega+\omega_3+\Lambda,\xi,\xi\p;-\omega_3,-\Lambda)\\
&-i\bII^\pm_{12}(\Omega,\omega,\xi,\xi\p;\Lambda,\omega_2)\Pi(\xi_{12},\xi\p+\omega_2+\Lambda)\bar{G}_{\bar{2}3}(\xi_{12},\xi\p+\omega_2+\Lambda,-\omega_2,\omega_3)\\
&\times\Pi(\xi_{13},\xi\p+\omega_3+\Lambda)\bar{G}_{\bar{3}\bar{1}}(\xi_{13},\xi\p+\omega_3+\Lambda,-\omega_3,-\Lambda)\,,
\end{aligned}
\label{eq:RG_Sigma}
\end{equation}
where the resolvent is given by
\begin{equation}
\Pi(E,\omega)=\frac{1}{E+i\omega-L_\D^\eff(E,\omega)}\,.
\end{equation}
Furthermore, on the r.h.s. we implicitly sum over repeated indices not occurring on the l.h.s., and integrate $\int_0^\Lambda\di\omega_2$ and $\int_0^\Lambda\di\omega_3$. 
Since except for $\Omega$ all frequencies are bound by $\Lambda$ and $\Lambda\rightarrow0$
during the RG flow, we expand around zero Matsubara frequency~\cite{PhysRevB.83.205103} $\bI^\pm_{11\p}(\Omega,\xi)=\bI_{11\p}^\pm(\Omega,0,\xi,0;0,0)$ and analogously for $\bII^\pm$. The frequency dependence of the vertices in lowest order is taken into account by setting all Matsubara frequencies on the r.h.s. of Eq.~\eqref{eq:RG_Ipm}-\eqref{eq:RG_IIpm} to zero and neglecting higher order contributions:
\begin{align}
\frac{\di }{\di\Lambda}\Big[\bI^\pm_{11\p}(\Omega,\omega,\xi,\xi\p;\omega_1,\omega_{1\p})-\bI^\pm_{11\p}(\Omega,\xi)\Big]
&=\begin{aligned}[t]&i\bI^\pm_{12}(\Omega,\xi)[\Pi(\xi_{12},\xi\p+\omega_1+\Lambda)-\Pi(\xi_{12},\Lambda)]\bar{G}_{\bar{2}1\p}(\xi_{12})\\
&+i\bar{G}_{12}(\Omega)[\Pi(\Omega_{12},\omega+\omega_1+\Lambda)-\Pi(\Omega_{12},\Lambda)] \bI^\pm_{\bar{2}1\p}(\Omega_{12},\xi)-(1\leftrightarrow 1\p)\,,
\end{aligned}\\
\frac{\di }{\di\Lambda}\Big[\bII^\pm_{11\p}(\Omega,\omega,\xi,\xi\p;\omega_1,\omega_{1\p})-\bII^\pm_{11\p}(\Omega,\xi)\Big]&=\begin{aligned}[t]&i\bI_{12}(\Omega)\big[\Pi(\Omega_{12},\omega+\omega_1+\Lambda)-\Pi(\Omega_{12},\Lambda)\big]\bI^\pm_{\bar{2}1\p}(\Omega_{12},\xi)\\
&\ +i\bII^\pm_{12}(\Omega,\xi)\big[\Pi(\xi_{12},\xi\p+\omega_1+\Lambda)-\Pi(\xi_{12},\Lambda)\big]\bar{G}_{\bar{2}1\p}(\xi_{12})-(1\leftrightarrow 1\p)\,.
\end{aligned}
\end{align}
Introducing the function $F(\Omega,\omega)$ defined by
\begin{equation}
i\Pi(\Omega,\omega)=\dd{}{\omega}F(\Omega,\omega)\,,
\end{equation}
 we can integrate these differential equations to
\begin{align}
\bI^\pm_{11\p}(\Omega,\omega,\omega_1,\omega_{1\p})&\cong \frac{1}{2}\bI^\pm_{11\p}(\Omega)
+\bI^\pm_{12}(\Omega)\left[F(\Omega_{12},\omega+\omega_1+\Lambda)-F(\Omega_{12},\Lambda)\right]\bG_{\21\p}(\Omega_{12})\nonumber\\
&\quad+\bG_{12}(\Omega)\left[F(\Omega_{12},\omega+\omega_1+\Lambda)-F(\Omega_{12},\Lambda)\right]\bI^\pm_{\21\p}(\Omega_{12})-(1\leftrightarrow1\p)\,,\\
\bII^\pm_{11\p}(\Omega,\omega;\omega_1,\omega_{1\p})&\cong \frac{1}{2}\bII^\pm_{11\p}(\Omega)+\bII^\pm_{12}(\Omega)\left[F(\Omega_{12},\omega+\omega_1+\Lambda)-F(\Omega_{12},\Lambda)\right]\bG_{\21\p}(\Omega_{12})\nonumber\\
&\quad+\bI_{12}(\Omega)\left[F(\Omega_{12},\omega+\omega_1+\Lambda)-F(\Omega_{12},\Lambda)\right]\bI^\pm_{\21\p}(\Omega_{12})-(1\leftrightarrow1\p)\,.
\end{align}
Here we used 
\begin{equation}
i\Pi(\Omega,\omega+\Lambda)\approx\ddl{} F(\Omega,\omega+\Lambda)\,,
\end{equation}
neglecting the implicit $\Lambda$-dependence due to $L_\D^\eff$ in the resolvents.
The analog expressions for the expanded vertex $\bG(\Omega)$ and $\bI(\Omega)$ can be found in Ref.~\onlinecite{PhysRevB.83.205103}. The RG equations~\eqref{eq:RG_Sigma}-\eqref{eq:RG_IIpm} thus reduce to
\begin{align}
\label{eq:RGeq_IB}
\ddl{\bI^\pm_{11\p}(\Omega,\xi)}
&=i\bI^\pm_{12}(\Omega,\xi)\Pi(\xi_{12},\Lambda)\bG_{\21\p}(\xi_{12})-i\bI^\pm_{1\p2}(\Omega,\xi)\Pi(\xi_{1\p2},\Lambda)\bG_{\21}(\xi_{1\p2})\nonumber\\
&\quad+i\bG_{12}(\Omega)\Pi(\Omega_{12},\Lambda)\bI^\pm_{\21\p}(\Omega_{12},\xi)-i\bG_{1\p2}(\Omega)\Pi(\Omega_{1\p2},\Lambda)\bI^\pm_{\21}(\Omega_{1\p2},\xi)\nonumber\\
&\quad+\bI^\pm_{23}(\Omega,\xi)\Pi(\xi_{23},\Lambda+\omega_3)\bG_{11\p}(\xi_{23})\Pi(\xi_{11\p32},\omega_3+\Lambda)\bG_{\bar{3}\2}(\xi_{11\p23})\nonumber\\
&\quad+\bG_{23}(\Omega)\Pi(\Omega_{23},\Lambda+\omega_3)\bI^\pm_{11\p}(\Omega_{23},\xi_{23})\Pi(\xi_{11\p32},\omega_3+\Lambda)\bG_{\bar{3}\2}(\xi_{11\p23})\nonumber\\
&\quad+\bG_{23}(\Omega)\Pi(\Omega_{23},\Lambda+\omega_3)\bG_{11\p}(\Omega_{23})\Pi(\Omega_{11\p32},\omega_3+\Lambda)\bI^\pm_{\bar{3}\2}(\Omega_{11\p23})\,,\\
\label{eq:RGeq_IAB}
\ddl{\bIAB_{11\p}^\pm(\Omega,\xi)}
&=i\bI_{12}(\Omega)\Pi(\Omega_{12},\Lambda)\bI^\pm_{\21\p}(\Omega_{12},\xi)-i\bI_{1\p2}(\Omega)\Pi(\Omega_{1\p2},\Lambda)\bI^\pm_{\21}(\Omega_{1\p2},\xi)\nonumber\\
&\quad+i\bIAB_{12}(\Omega,\xi)\Pi(\xi_{12},\Lambda)\bG_{\21\p}(\xi_{12})-i\bIAB_{1\p2}(\Omega,\xi)\Pi(\xi_{1\p2},\Lambda)\bG_{\21}(\xi_{1\p2})\nonumber\\
&\quad+\bI_{23}(\Omega)\Pi(\Omega_{23},\Lambda+\omega_3)\bI^\pm_{11\p}(\Omega_{23},\xi)\Pi(\xi_{11\p32},\omega_3+\Lambda)\bG_{\bar{3}\2}(\xi_{11\p23})\nonumber\\
&\quad+\bI_{23}(\Omega)\Pi(\Omega_{23},\Lambda+\omega_3)\bG_{11\p}(\Omega_{23})\Pi(\Omega_{11\p32},\omega_3+\Lambda)\bI^\pm_{\bar{3}\2}(\Omega_{11\p23},\xi)\nonumber\\
&\quad+\bIAB_{23}(\Omega,\xi)\Pi(\xi_{23},\Lambda+\omega_3)\bG_{11\p}(\xi_{23})\Pi(\xi_{11\p32},\omega_3+\Lambda)\bG_{\bar{3}\2}(\xi_{11\p23}),\\
\label{eq:RGeq_SigmaII}
\ddl{\Sigma_{\II}^\pm(\Omega,\xi)}&=-i\bI_{12}(\Omega)K(\Omega_{12})\bI^\pm_{\2\bar{1}}(\Omega_{12},\xi)-i\bIAB_{12}(\Omega,\xi)K(\xi_{12})\bG_{\2\1}(\xi_{12})\nonumber\\
&\quad-2i\bI_{12}(\Omega)K(\Omega_{12})\bI^\pm_{\23}(\Omega_{12},\xi)K(\xi_{\23})\bG_{\3\1}(\xi_{\23})\nonumber\\
&\quad-2i\bI_{12}(\Omega)K(\Omega_{12})\bG_{\23}(\Omega_{12})K(\Omega_{13})\bI^\pm_{\3\1}(\Omega_{13},\xi)\nonumber\\
&\quad-2i\bIAB_{12}(\Omega,\xi)K(\xi_{12})\bG_{\23}(\xi_{12})K(\xi_{13})\bG_{\3\1}(\xi_{13})\,,
\end{align}
where we used 
\begin{equation}
K(\Omega)=i\int\limits_0^\Lambda\di\omega\Pi(\Omega,\omega+\Lambda)\,.
\label{eq:K}
\end{equation}

\emph{Weak-coupling analysis above $\Lambda_c$.}
As discussed in detail in Refs.~\onlinecite{EPJST} and \onlinecite{PhysRevB.80.045117} for $\Lambda>\Lambda_c=\max\{|\Omega|,|V|,|h|\}$ the cutoff scales in the resolvents can be neglected. This leads to a reference solution $\bG^{(1)}\propto J$, which can be used as a starting point for a systematic expansion of the RG equations in orders of the coupling constant $J$ at scale $\Lambda$.

For the solution of the flow equations it is important to note that  terms proportional to $J^n/\Lambda$ on the r.h.s. lead to contributions $J^{n-1}$. Furthermore, by expanding the resolvent it can be shown\cite{EPJST,PhysRevB.80.075120,PhysRevB.80.045117}, that the vertices and the corresponding RG equations can be split in a frequency dependent and a frequency independent part
\begin{equation}
\bI^\pm(\Omega,\xi)=\bI^{\pm(1)}+i\bI^{\pm(2a_1)}+\bI^{\pm(2a_2)}+\bI^{\pm(2b)}(\Omega,\xi)\,.
\end{equation}
Here the superscripts $(1)$ and $(2)$ indicate the order of the vertex in $J$. For the frequency-independent part of the RG equations we consider a differential equation for the imaginary part $(2a_1)$ and one for the real part containing the sum of $\bI^{\pm(1)}$ and $\bI^{\pm(2a_2)}$ of the vertices 
\begin{subequations}
\label{RG_I}
\begin{align}
\label{eq:RG_Ipm_2}
\dd{\bI_{11\p}^{\pm(1+2a_2)}}{\Lambda}
&=\frac{1}{\Lambda}\Big[\bI^\pm_{12}\bG^{(1)}_{\21\p}-\bI^\pm_{1\p2}\bG^{(1)}_{\21}+\bG^{(1)}_{12}\bI^\pm_{\21\p}-\bG^{(1)}_{1\p2}\bI^\pm_{\21}\nonumber\\
&\quad+\bI^\pm_{12}\bG^{(2a_2)}_{\21\p}-\bI^\pm_{1\p2}\bG^{(2a_2)}_{\21}+\bG^{(2a_2)}_{12}\bI^\pm_{\21\p}-\bG^{(2a_2)}_{1\p2}\bI^\pm_{\21}\nonumber \\
&\quad-\bI_{12}^{\pm(1)}Z^{(1)}\bG_{\21\p}^{(1)}-\bG_{12}^{(1)}Z^{(1)}\bI_{\21\p}^{\pm(1)}+\bI_{1\p2}^{\pm(1)}Z^{(1)}\bG_{\21}^{(1)}+\bG_{1\p2}^{(1)}Z^{(1)}\bI_{\21}^{\pm(1)}\nonumber\\
&\quad-\frac{1}{2}\bI_{23}^{\pm(1)}\bG_{11\p}^{(1)}\bG_{\3\2}^{(1)}-\frac{1}{2}\bG_{23}^{(1)}\bI_{11\p}^{\pm(1)}\bG_{\3\2}^{(1)}-\frac{1}{2}\bG_{23}^{(1)}\bG_{11\p}^{(1)}\bI_{\3\2}^{\pm(1)}
\Big],\\
\label{eq:RG_Ipm_2a1}
	\ddl{\bI^{\pm(2a_1)}_{11\p}}&=\frac{1}{\Lambda}\Big[\bI^{\pm(1)}_{12}\bG^{(2a_1)}_{\21\p}+\bI^{\pm(2a_1)}_{12}\bG^{(1)}_{\21\p}-\bI^{\pm(1)}_{1\p2}\bG^{(2a_1)}_{\21}-\bI^{\pm(2a_1)}_{1\p2}\bG^{(1)}_{\21}\nonumber\\
	&\quad+\bG^{(1)}_{12}\bI^{\pm(2a_1)}_{\21\p}+\bG^{(2a_1)}_{12}\bI^{\pm(1)}_{\21\p}-\bG^{(1)}_{12}\bI^{\pm(2a_1)}_{\21\p}-\bG^{(2a_1)}_{12}\bI^{\pm(1)}_{\21\p}
	\Big],\\
	\label{eq:RG_IABpm}
	\ddl{\bIAB^{\pm(1+2a_2)}_{11\p}}&=\frac{1}{\Lambda}\Big[\bI^{(1)}_{12}\bI^{\pm(1)}_{\21\p}-\bI^{(1)}_{1\p2}\bI^{\pm(1)}_{\21}+\bI^{(1)}_{12}\bI^{\pm(2)}_{\21\p}-\bI^{(1)}_{1\p2}\bI^{\pm(2)}_{\21}+\bIAB^\pm_{12}\bG^{(1)}_{\21\p}-\bIAB^\pm_{1\p2}\bG^{(1)}_{\21}\nonumber\\
&\quad-\bI^{(1)}_{12}Z^{(1)}\bI^{\pm(1)}_{\21\p}+\bI^{(1)}_{1\p2}Z^{(1)}\bI^{\pm(1)}_{\21}-\bIAB^{\pm(1)}_{12}Z^{(1)}\bG^{(1)}_{\21\p}+\bIAB^{\pm(1)}_{1\p2}Z^{(1)}\bG^{(1)}_{\21}\nonumber\\
&\quad-\frac{1}{2}(\bI^{(1)}_{23}\bI^{\pm(1)}_{11\p}\bG^{(1)}_{\3\2}+\bI^{(1)}_{23}\bG^{(1)}_{11\p}\bI^{\pm(1)}_{\3\2}+\bIAB^{\pm(1)}_{23}\bG^{(1)}_{11\p}\bG^{(1)}_{\3\2})
\Big],\\
\label{eq:RG_IABpm_2a1}
\ddl{\bIAB^{\pm(2a_1)}_{11\p}}&=\frac{1}{\Lambda}\Big[\bI^{(1)}_{12}\bI^{\pm(2a_1)}_{\21\p}+\bI^{(2a_1)}_{12}\bI^{\pm(1)}_{\21\p}-\bI^{(1)}_{1\p2}\bI^{\pm(2a_1)}_{\21}-\bI^{(2a_1)}_{1\p2}\bI^{\pm(1)}_{\21}\nonumber\\
&\quad+\bIAB^{\pm(1)}_{12}\bG^{(2a_1)}_{\21\p}+\bIAB^{\pm(2a_1)}_{12}\bG^{(1)}_{\21\p}-\bIAB^{\pm(1)}_{1\p2}\bG^{(2a_1)}_{\21}-\bIAB^{\pm(2a_1)}_{1\p2}\bG^{(1)}_{\21}
\Big]\,,
\end{align}
\end{subequations}
where $Z^{(1)}$ parametrizes the frequency dependence of the Liouvillian in first order~\cite{PhysRevB.80.045117} by $L_\D^{(1)}(\Omega)=L_\D^{(1)}-\Omega Z^{(1)}$.
The initial conditions of the RG equations are given by the discrete RG step \eqref{eq:ic_IB}-\eqref{eq:ic_IAB}. We note that only the imaginary parts of $\bI^{\pm a}$ and $\bIAB^{\pm a}$ are generated during the discrete step. Thus the real part of $\bI^\pm$ is given by the bare vertex; $\bIAB^\pm$ is initially zero.

The frequency-dependent part $\bI^{\pm}$ and $\bIAB^{\pm}$ can be integrated to
\begin{align}
	\bI_{11\p}^{\pm(2b)}(\Omega,\xi)&=\bI_{12}^{\pm(1)}\ln{\frac{\Lambda-i\xi_{12}+iL_\D^{(0)}}{\Lambda}}
\bG_{\21'}^{(1)}+\bG_{12}^{(1)}\ln{\frac{\Lambda-i\Omega_{12}+iL_\D^{(0)}}{\Lambda}}
\bI_{\21'}^{\pm(1)}-(1\leftrightarrow1\p),\\
\bIAB_{11\p}^{\pm(2b)}(\Omega,\xi)&=\bIAB_{12}^{\pm(1)}\ln{\frac{\Lambda-i\xi_{12}+iL_\D^{(0)}}{\Lambda}}
\bG_{\21'}^{(1)}+\bI_{12}^{(1)}\ln{\frac{\Lambda-i\Omega_{12}+iL_\D^{(0)}}{\Lambda}}
\bI_{\21'}^{\pm(1)}-(1\leftrightarrow1\p)\,.
\end{align}

In order to distinguish between the different orders in $J$ we expand $K(z)$ as
\begin{equation}
K_{\Lambda}(z)=\ln\frac{2\Lambda-iz}{\Lambda-iz}=\frac{iz}{2\Lambda}+\wt{K}_\Lambda(z)\,.
\end{equation}
The terms proportional to $J^n/\Lambda$  on the r.h.s. lead to contributions $J^{n-1}$ after the integration over $\Lambda$, while those proportional to $\wt{K}_\Lambda(z)$ remain of the same order.

Similarly to the vertices, the RG equation of the kernel $\Sigma_{\II}^\pm$ can be split in one for the real and one for the imaginary part. Since for the derivation of the noise only the imaginary part of $\Sigma_{\IAB}^\pm$ is needed, we will restrict our analysis to $\Im\Sigma_{\IAB}^\pm$ here. The real part will be considered in Appendix~\ref{sssec:acG}, for the calculation of the imaginary part of the ac conductance. In analogy to the equations of $\bI^\pm$ and $\bIAB^\pm$ the imaginary part of $\Sigma_{\II}^\pm$ is given by
\begin{equation}
\begin{aligned}[b]
\ddl{\Im\Sigma_{\II}^\pm}=&\frac{1}{2\Lambda}\Big[\bI_{12}^{(1)}(\Omega_{12}-L_\D^{(0)})\bI^{\pm(2a_1)}_{\2\1}\!+\bI_{12}^{(2a_1)}(\Omega_{12}-L_\D^{(0)})\bI^{\pm(1)}_{\2\1}\!+\bIAB^{\pm(1)}_{12}(\xi_{12}-L_\D^{(0)})\bG_{\2\1}^{(2a_1)}\!+\bIAB^{\pm(2a_1)}_{12}(\xi_{12}-L_\D^{(0)})\bG_{\2\1}^{(1)}\Big]\\
&-\bI_{12}^{(1)}\Im\wtK(\Omega_{12})\bI^{\pm(1)}_{\2\1}-\bIAB^{\pm(1)}_{12}\Im\wtK(\xi_{12})\bG_{\2\1}^{(1)}\,,
\end{aligned}
\label{eq:RG_ImSigma}
\end{equation}
\end{widetext}
with initial condition given by Eq.~\eqref{eq:ic_Sigma}. 
In order to distinguish between contributions involving $\wt{K}_\Lambda(z)$ and the ones proportional to $1/\Lambda$ it is useful to split the RG equation for $\Sigma_{\II}^\pm$ in 
\begin{align}
\label{eq:ImSigma_2a}
	&\begin{aligned}[b]
	\ddl{\Im\Sigma_{\II}^{\pm(2a)}}=&-\bI_{12}^{(1)}\Im\wtK(\Omega_{12})\bI^{\pm(1)}_{\2\1}\\
	&-\bIAB^{\pm(1)}_{12}\Im\wtK(\xi_{12})\bG_{\2\1}^{(1)} ,
	\end{aligned}\\
	\label{eq:ImSigma_2b}
	&\begin{aligned}[b]
	\ddl{\Im\Sigma_{\II}^{\pm(2b)}}=&\frac{1}{2\Lambda}\Big[\bI_{12}^{(1)}(\Omega_{12}-L_\D^{(0)})\bI^{\pm(2a_1)}_{\2\1}\\
	&+\bI_{12}^{(2a_1)}(\Omega_{12}-L_\D^{(0)})\bI^{\pm(1)}_{\2\1}\\
&+\bIAB^{\pm(1)}_{12}(\xi_{12}-L_\D^{(0)})\bG_{\2\1}^{(2a_1)}
\\
&+\bIAB^{\pm(2a_1)}_{12}(\xi_{12}-L_\D^{(0)})\bG_{\2\1}^{(1)}\Big]\,.
	\end{aligned}
\end{align}
Using $\wtK(z)=\di\wt{F}_\Lambda(z)/\di\Lambda$, with
\begin{equation}
\wt{F}_\Lambda(z)=\Lambda\ln\frac{2\Lambda-iz}{\Lambda-iz}-\frac{iz}{2}\left(\ln\frac{(2\Lambda-iz)\Lambda}{2(\Lambda-iz)^2}+1\right)\,,
\label{eq:K_F}
\end{equation}
\eqref{eq:ImSigma_2a} can be integrated to
\begin{equation}
\begin{aligned}
\Im\Sigma_{\II}^{\pm(2a)}&&\\
&&\hspace{-1cm}=-\bI_{12}^{(1)}\Im\wt{F}_{\Lambda_c}(\Omega_{12})\bI^{\pm(1)}_{\2\1}\!-\bIAB^{\pm(1)}_{12}\Im\wt{F}_{\Lambda_c}(\xi_{12})\bG_{\2\1}^{(1)}\,,
\end{aligned}
\end{equation}
where we used $\wt{F}_\Lambda(z)\rightarrow\Lambda\ln2+\Lambda\mc{O}(z/\Lambda)^2$ for $\Lambda\gg|z|$. Thus the contribution at $\Lambda_0$ is canceled by the first term of the initial condition \eqref{eq:ic_Sigma} for~\cite{PhysRevB.80.045117}
\begin{equation}
\Lambda_0=\frac{\pi^2}{16\ln2}D\,.
\end{equation}

\emph{Weak-coupling analysis below $\Lambda_c$.}
As explained in Ref.~\onlinecite{PhysRevB.80.045117}, up to $\Lambda_c$ we resummed all leading and subleading logarithmic contributions in $\ln(D/\Lambda_c)$ for the renormalized vertex. Thus at $\Lambda=\Lambda_c$ the bare coupling constant is replaced by t
\begin{equation}
J_c=\frac{1}{2\ln\frac{\Lambda_c}{T_K}}\, ,
\label{eq:J_c}
\end{equation}
providing the starting point for an expansion in $J_c\ll1$ for $\Lambda_c\gg T_K$.At the same time the Liouvillian in the resolvents is replaced by the full effective one $L_\D^\eff(z)$.

Since we stop the flow of the coupling $J$ at $\Lambda_c$ also the RG flow of all vertices is stopped at this scale. The current-current kernel $\Sigma_{\II}^\pm$ at $\Lambda=0$ is calculated perturbatively in $J_c$ by replacing all vertices by their values at $\Lambda=\Lambda_c$ in the following indicated by the index $c$ (e.g. $\bG^{c(1)}$). Carrying out this replacement in Eq.~\eqref{eq:RGeq_SigmaII} we find for the RG equation of the imaginary part of the current-current kernel up to second order in $J_c$
\begin{equation}
\begin{aligned}
\ddl{\Im\Sigma_{\II}^{\pm(2)}}&&\\
&&\hspace{-1.1cm}=-\bI_{12}^{c(1)}\Im K(\Omega_{12})\bI^{c\pm(1)}_{\2\1}\!-\bIAB^{c\pm(1)}_{12}\Im K(\xi_{12})\bG_{\2\1}^{c(1)}\,,
\end{aligned}
\label{eq:ImSigma_bLc}
\end{equation}
which can be easily integrated from $\Lambda_c$ to $0$ by using $K(z)=\di F_\Lambda(z)/\di\Lambda$ with
\begin{equation}
F_\Lambda(z)=\wt{F}_\Lambda(z)+\frac{iz}{2}\left(\ln\frac{i\Lambda}{2z}+1\right)\,.
\end{equation}
The contribution proportional to $\wt{F}_{\Lambda_c}(z)$ is canceled by the corresponding term from above $\Lambda_c$, and hence we find 
\begin{equation}
\begin{aligned}[b]
\Im\Sigma_{\II}^\pm=&-\bI_{12}^{c(1)}\Im\frac{\Omega_{12}-L_\D^\eff(\Omega)}{2}\\
&\qquad\times\left(\ln\frac{i\Lambda_c}{2(\Omega_{12}-L_\D^\eff(\Omega))}+1\right)\bI^{c\pm(1)}_{\2\1}\\
&-\bIAB^{c\pm(1)}_{12}\Im\frac{\xi_{12}-L_\D^\eff(\xi)}{2}\\
&\qquad\times\left(\ln\frac{i\Lambda_c}{2(\xi_{12}-L_\D^\eff(\xi))}+1\right)\bG_{\2\1}^{c(1)}\,,
\end{aligned}
\end{equation}
where $\wt{F}_{\Lambda=0}=0$.

In order to obtain analytic solutions for the RG equations we decompose the effective Liouvillian into eigenvalues and projectors $L_\D^\eff(z)=\sum_i\lambda_i(z)P_i(z)$ and expand around the poles  given by the self-consistency equations $z_i=\lambda_i(z_i)$. 
 With this approximation $\Im\Sigma_{\II}^\pm$ at $\Lambda=0$ is given by
\begin{equation}
\begin{aligned}
&\Im\Sigma_{\II}^\pm\\
&=-\sum_{i}\frac{\Omega_{12}-\Re z_i}{2}\arctan\frac{\Omega_{12}-\Re z_i}{\Im z_i}\bI_{12}^{c(1)}P_i\bI^{c\pm(1)}_{\2\1}\\
&\quad-\sum_{i}\frac{\xi_{12}-\Re z_i}{2}\arctan\frac{\xi_{12}-\Re z_i}{\Im z_i}\bIAB^{c\pm(1)}_{12}P_i\bG_{\2\1}^{c(1)}\,.
\end{aligned}
\label{eq:ImSigma_bLc_p}
\end{equation}

As explained in detail in Ref.~\onlinecite{EPJST} the eigenvalue $z_i=0$ characterizing the stationary state, could in principle lead to divergencies. Performing the discrete RG step its contribution vanishes, if the vertex $\bG$ stands right of the projector $P_0$ since $P_0\bG=0$. This is the case for the second contribution of \eqref{eq:ImSigma_bLc_p}, but not for the first one.
From the definitions \eqref{eq:def_I} we obtain the relations $\bI^+_{11\p}=2\wt{G}_{11\p}$ and $\bI^-_{11\p}=2\bG_{11\p}$. Since $\bI^-$ is proportional to $\bG$, also the contribution of $P_0$ in the first term of \eqref{eq:ImSigma_bLc_p} is zero. For $\bI^+$ this does not hold and the zero eigenvalue will lead to a contribution characterized by the absence of any finite relaxation rate leading to a sharp kink in the symmetric noise and a discontinuity in its derivative. 
However, there are no contributions of the zero eigenvalue~\cite{thesis} in order $J_c^3\ln(\Lambda_c/|x+i\Gamma|)$ and thus no divergent logarithmic contributions emerge in $\Im\Sigma_{\II}^+$.

\subsubsection{Application to the isotropic Kondo model}\label{sssec:isoKondo}

In the following we solve the RG equations set up in the previous section explicitly for the isotropic Kondo model. 
The interaction part of the Hamiltonian~\eqref{eq:model} is given by
	\begin{equation}
	V=\frac{1}{2}g_{11\p}:a_1a_{1\p}:\,,
	\end{equation}
	where we used the notation $1=\eta\alpha\s\omega$ and sum (integrate) implicitly over all indices (frequencies). Here
	$g_{11\p}$ is the coupling vertex acting on the dot states only and :.: denotes normal-ordering of the reservoir field operators, meaning that no contractions within the normal-ordered product are allowed. A contraction is defined by
\begin{equation}
\label{contraction}
{a_1\,a_{1\p}
  \begin{picture}(-20,11) 
    \put(-22,8){\line(0,1){3}} 
    \put(-22,11){\line(1,0){12}} 
    \put(-10,8){\line(0,1){3}}
  \end{picture}
  \begin{picture}(20,11) 
  \end{picture}
}
\,\equiv\,
\Ew{a_1 a_{1\p}}_{\rho_{\res}}=\delta_{1\1\p}\rho(\omega)f_\alpha(\eta\omega)\,,
\end{equation}
with $f_\alpha(\eta\omega)=(e^{\omega/T_\alpha}+1)^{-1}=1-f_\alpha(-\omega)$ the Fermi distribution function at the corresponding temperature $T_\alpha$ of the reservoir and $\delta_{11\p}\equiv \delta_{\eta\eta\p}\delta_{\alpha\alpha\p}\delta_{\s\s\p}\delta(\omega-\omega\p)$ is the $\delta$ distribution in compact notation. Furthermore, we introduce the cutoff band width $D$ via the density of states
\begin{equation}
\rho(\omega)=\frac{D^2}{D^2+\omega^2}\,.
\end{equation}
	
	For the isotropic Kondo model we consider (see Fig.~\ref{fig:fig1} and Eq.~\eqref{eq:model}) the coupling vertex 
	\begin{equation}
	g_{11\p}=\frac{1}{2}\begin{cases}(J_{\alpha\alpha\p})_0S^i\s_{\s\s\p}^i &\text{for }\eta=-\eta\p=+\\
	-(J_{\alpha\p\alpha})_0S^i\s_{\s\p\s}^i &\text{for }\eta=-\eta\p=-
			\end{cases}\,.
	\label{eq:vertex_g}
	\end{equation}
In Liouville space it reads
\begin{equation}
G^{pp}_{11\p}=\frac{1}{2}\begin{cases}(J_{\alpha\alpha\p})_0L^{pi}\s_{\s\s\p}^i&\text{for }\eta=-\eta\p=+\\
-(J_{\alpha\p\alpha})_0L^{pi}\s_{\s\p\s}^i&\text{for }\eta=-\eta\p=-\,,
\end{cases}
\label{eq:vertex_G}
\end{equation}
with the spin superoperators $\vec{L}^p=(L^{px},L^{py},L^{pz})$ defined by their action on an arbitrary operator $A$ in the dot Hilbert space
\begin{equation}
\vec{L}^+A=\vec{S}A\,,\quad \vec{L}^-A=-A\vec{S}\,.
\end{equation}
The explicit matrix structure of these spin matrices can be found in Ref.~\onlinecite{PhysRevB.80.075120}. For the solution of the RG equations these  matrices will always occur in the combinations
\begin{subequations}
\begin{align}
	\vec{L}^1&=\frac{1}{2}(\vec{L}^+-\vec{L}^-)-i\vec{L}^+\times\vec{L}^-\\
	\vec{L}^2&=-\frac{1}{2}(\vec{L}^++\vec{L}^-)\\
	\vec{L}^3&=\frac{1}{2}(\vec{L}^+-\vec{L}^-)+i\vec{L}^+\times\vec{L}^-\\
	L^a&=\frac{3}{4}\E+\vec{L}^+\cdot\vec{L}^-\\
	L^b&=\frac{1}{4}\E-\vec{L}^+\cdot\vec{L}^-\\
	L^c&=\frac{1}{2}\E+2L^{+z}L^{-z}\\
	L^h&=L^{+z}+L^{-z}=-2L^{2z}\,.
\end{align}
\end{subequations}
With these spin superoperators the Liouvillian in zeroth order is given by
\begin{equation}
L_\D^{(0)}=\kom{H_\D}{\cdot}=h_0L^h\,.
\end{equation} 

 The results for the coupling vertex, the Liouvillian and the current vertex $I$ for $\eta=-\eta\p=+$ determined in Refs.~\onlinecite{PhysRevB.80.045117,EPJST} read
\begin{subequations}
\label{eq:vertices_iK}
\begin{align}
	\bG_{11\p}^{(1)}&=-J_{\alpha\alpha\p}\vec{L}^2\vec{\s}_{\s\s\p}\\
	\wt{G}_{11\p}^{(1)}&=\frac{1}{2}J_{\alpha\alpha\p}(\vec{L}^1+\vec{L}^3)\vec{\s}_{\s\s\p}\\
	\bG_{11\p}^{(2a_1)}&=\frac{\pi}{2}J_{\alpha\beta}J_{\beta\alpha\p}\vec{L}^3\vec{\s}_{\s\s\p}\\
	L_\D^{(1)}&=\frac{1}{2}\tr Jh_0L^h\\
	\label{eq:I_iK}
\bI^{(1)}_{11\p}&=\frac{1}{2}J_{\alpha\alpha\p}^L \vec{L}^{1}\vec{\s}_{\s\s\p}\\
\bI^{(2a_1)}_{11\p}&=-\frac{3\pi}{8}(J_{\alpha\beta}^L J_{\beta\alpha\p}-J_{\alpha\beta} J_{\beta\alpha\p}^L)L^b\delta_{\s\s\p}\,,
\end{align}
\end{subequations}
where $\vec{\s}$ is a vector consisting of the Pauli matrices and $J_{\alpha\alpha\p}^L=c_{\alpha\alpha\p}^LJ_{\alpha\alpha\p}$. The vertex for $\eta=-\eta\p=-$ is obtained by using $\bG_{11\p}=-\bG_{1\p 1}$, which also holds for the current vertices. Furthermore the coupling constant $J_{\alpha\alpha\p}=2\sqrt{x_\alpha x_{\alpha\p}}J$ fulfills the poor man's scaling equation 
\begin{equation}
 \ddl{J(\Lambda)}=-\frac{2}{\Lambda}J(\Lambda)^2
\label{eq:PMS}
\end{equation}
which is solved by
\begin{equation}
J(\Lambda)=\frac{1}{2\ln\frac{\Lambda}{T_K}}\,,\quad T_K=\Lambda_0e^{-1/(2J_0)}\,,
\label{eq:pms}
\end{equation}
where $J_0=J(\Lambda_0)$ is the initial value of the coupling constant at $\Lambda_0$.

According to Eqs.~\eqref{eq:Ip} and \eqref{eq:Im}, the initial values for $\bI^\pm$ and $\wt{I}^\pm$ are given by
\begin{subequations}
\label{eq:Ipm_iK}
\begin{align}
\label{eq:bIp}
	\bI_{11\p}^{+(1)}&=2\wt{G}_{11\p}^{(1)}=J_{\alpha\alpha\p}^L(\vec{L}^1+\vec{L}^3)\vec{\s}_{\s\s\p}\\
	\label{eq:wtIp}
	\wt{I}_{11\p}^{+(1)}&=2\bG_{11\p}^{(1)}=-2J_{\alpha\alpha\p}^L\vec{L}^2\vec{\s}_{\s\s\p}\\
	\label{eq:bIm}
	\bI_{11\p}^{-(1)}&=2\bG_{11\p}^{(1)}=-2J_{\alpha\alpha\p}^L\vec{L}^2\vec{\s}_{\s\s\p}\\
	\label{eq:wtIm}
	\wt{I}_{11\p}^{-(1)}&=2\wt{G}_{11\p}^{(1)}=J_{\alpha\alpha\p}^L(\vec{L}^1+\vec{L}^3)\vec{\s}_{\s\s\p}\,.
\end{align}
\end{subequations}
Inserting these initial vertices in Eqs.~\eqref{RG_I} and using the poor man's scaling equation \eqref{eq:PMS} the vertices $\bI_{11\p}^\pm$ as well as $\bIAB_{11\p}^\pm$ can be determined up to second order in $J$. Since for the derivation of the imaginary part of the current-current kernel the real part of the vertices is needed only in first order, we skip the second order of the real part here. In first order $I^\pm$ is given by its initial value with $\Lambda$-dependent coupling $J$. The imaginary part of the second order is given by its initial condition \eqref{eq:ic_IB} with the initial couplings $J_0$ replaced by the $\Lambda$-dependent ones
\begin{subequations}
\begin{align}
&\begin{aligned}\bI^{+(2a_1)}_{11\p}=
	&-i\frac{3\pi}{4}\big(J_{\alpha\beta}^L J_{\beta\alpha\p}-J_{\alpha\beta}J_{\beta\alpha\p}^L\big)L^b\delta_{\s\s\p}\\
	&+i\frac{\pi}{4}\big(J_{\alpha\beta}^L J_{\beta\alpha\p}-J_{\alpha\beta}J_{\beta\alpha\p}^L\big)L^a\delta_{\s\s\p}\\
	&-i\pi(J_{\alpha\beta}^L J_{\beta\alpha\p}+J_{\alpha\beta}J_{\beta\alpha\p}^L)L^{2i}\s_{\s\s\p}^i\,.
	\end{aligned}\\
	&\bI^{-(2a_1)}_{11\p}=i\pi(J_{\alpha\beta}J_{\beta\alpha\p}^L+J_{\beta\alpha\p}J_{\alpha\beta}^L)L^{3i}\s_{\s\s\p}^i
\end{align}
\end{subequations}
The generated vertex $\bIAB^\pm$ is initially given by \eqref{eq:ic_IAB}
\begin{align}
\bIAB_{11\p}^{+a}&=i\pi J_{\alpha\beta}^LJ_{\beta\alpha\p}^L L^{1i}\s_{\s\s\p}^i\\
\bIAB_{11\p}^{-a}&=0\,
\end{align}
with no contribution in first order. Solving Eqs.~\eqref{eq:RG_IABpm} and \eqref{eq:RG_IABpm_2a1} we find that $\bIAB^+$ remains zero in first order, while $\bIAB^-$ is given by
\begin{equation}
\label{eq60}
\bIAB^{-(1)}_{11\p}=a_{\alpha\alpha\p}^{LL}L^{1i}\s_{\s\s\p}^i\,,
\end{equation}
with
\begin{equation}
	a_{\alpha\alpha\p}^{LL}=\begin{aligned}[t]&\frac{1}{4}\Big[-\frac{1}{2}\frac{J^2}{J_0}+\frac{1}{2}J_0\Big]\E+\frac{1}{4}\Big[J-\frac{1}{2}\frac{J^2}{J_0}-\frac{1}{2}J_0\Big]\s^x\,.\end{aligned}
\end{equation}
We note that the contribution of~\eqref{eq60} to the antisymmetric noise as well as to the ac conductance will turn out to vanish.
For the contributions in second order $\bIAB^{-}$ is zero while of $\bIAB^+$ is given by
\begin{align}
	\bIAB_{11\p}^{(2a_1)}&=\frac{\pi}{2}(J_{\alpha\beta}^L J_{\beta\alpha\p}^L+J_{\beta\alpha\p}^L J_{\alpha\beta}^L)\vec{L}^1\vec{\s}_{\s\s\p}\,.
\end{align}
The RG equation~\eqref{eq:ImSigma_2b} for the contribution from $\Lambda>\Lambda_c$ of $\Sigma_\II^\pm(\Omega,\xi)$ can be solved by inserting these vertices and using the poor man's scaling equation~\eqref{eq:PMS}.

Using for the poles $z_i$ and projectors $P_i$ (see Ref.~\onlinecite{PhysRevB.80.045117})
\begin{align}
\label{eq:poles}
	&z_0=0&\qquad&P_0=L^b+2ML^{3z}&\\
	&z_1=-i\Gamma_1&\qquad&P_1=L^a-L^c-2ML^{3z}&\\
	&z_\pm=\pm h-i\Gamma_2&\qquad&P_\pm=\frac{1}{2}(L^c\pm L^h)\,,&
\end{align}
with the renormalized magnetic field $h$, relaxation rates $\Gamma_{1/2}$ and the magnetization $M$, Eq.~\eqref{eq:ImSigma_bLc_p} 
gives for the current-current kernel in the limit $\xi \to 0^+$,
\begin{align}
&\begin{aligned}
	\Im\Sigma_{\II}^+=&\frac{\pi}{2}J_\nd^2h L^{1z}+\frac{\pi}{8}J_\nd^2\sum_{\alpha,\s=\pm}|\Omega+\alpha V+\s h|_2L^b\\
	&+\frac{\pi}{4}J_\nd^2\sum_{\alpha=\pm}\Big[|\Omega+\alpha V|-|\Omega-\alpha V|_1\Big]ML^{1z}\\
	&+\frac{\pi}{2}J_\nd^2\sum_{\alpha=\pm}|\Omega+\alpha V|_1L^b
	\end{aligned}\\
	&\Im\Sigma_{\II}^-=\frac{3\pi}{4}J_\nd^2\Omega L^b+\frac{\pi}{8}J_\nd^2\sum_{\alpha,\s=\pm}\s|\Omega+\alpha V+\s h|_2 L^{1z}\,.
\end{align}
The symmetric and the antisymmetric noise~\eqref{eq:S} are obtained by multiplying with the stationary density matrix~\cite{PhysRevB.80.045117} $\rho_\st=\frac{1}{2}\E+2MS^z$ from the right, and performing the trace over the dot degrees of freedom, with $\Tr_\D L^b\rho_\st=1$ and $\Tr_\D L^{1z}\rho_\st=2M$. 

\subsubsection{AC conductance}\label{sssec:acG}

The real part of the ac conductance is given by Eq.~\eqref{eq:ReG_CCC} and thus can directly be calculated from the antisymmetric current noise \eqref{eq:Sm_final}.  

According to Eq.~\eqref{eq:G}, the imaginary part of the ac conductance is given by
\begin{equation}
\Im G(\Omega)=\frac{1}{\Omega}\big[\Im C^-(\Omega)-\Im C^-(0)\big]\,,
\label{eq:ImG}
\end{equation}
with
\begin{equation}
\Im C^-(\Omega)=-\Tr_\D[\Re\Sigma_{\IAB}^-(\Omega,i0^+)\rho_\D^\st]\,.
\label{eq:ImC}
\end{equation}
Starting from Eq.~\eqref{eq:RGeq_SigmaII}, and following the calculation of $\Im\Sigma_{\IAB}^-$ we obtain the flow equation for the real part
\begin{equation}
\begin{aligned}[b]
&\ddl{\Re\Sigma_{\IAB}^-}\\
&=\frac{1}{2\Lambda}\bI_{12}^{(1)}(\Omega_{12}\!-\!L_\D^{(0)})\bI^{-(2a_2)}_{\2\1}\!+\!\frac{1}{2\Lambda}\bI_{12}^{(2a_2)}(\Omega_{12}\!-\!L_\D^{(0)})\bI^{-(1)}_{\2\1}\\
&\ \ +\frac{1}{2\Lambda}\bIAB^{(1)}_{12}(\xi_{12}\!-\!L_\D^{(0)})\bG_{\2\1}^{(2a_2)}\!+\!\frac{1}{2\Lambda}\bIAB^{(2a_2)}_{12}(\xi_{12}\!-\!L_\D^{(0)})\bG_{\2\1}^{(1)}\\
&\ \ -\bI_{12}^{(1)}\Re\wtK(\Omega_{12})\bI^{-(1)}_{\2\1}-\bIAB^{(1)}_{12}\Re\wtK(\xi_{12})\bG_{\2\1}^{(1)}\,.
\end{aligned}
\end{equation}
As a consequence of \eqref{eq:ImG}, all $\Omega$-independent terms will not contribute. 
In addition the current vertices $\bI^{(2a_2)}$ and $\bI^{-(2a_2)}$ determined by the RG equation \eqref{eq:RG_Ipm_2} do not contribute due to the matrix structure as $\bI^{(2a_2)}\propto \vec{L}^{1}$ and $\bI^{-(2a_2)}\propto \vec{L}^{1}$.
Using \eqref{eq:I_iK} and \eqref{eq:bIm} for $\bI^{(1)}$ and $\bI^{-(1)}$, we thus obtain
\begin{equation}
\Re\Sigma_{\II}^-=\frac{1}{4}J_\nd^2\sum_{\alpha,\s=\pm}\s\mc{L}_2(\Omega+\alpha V+\s h)L^{1z}\,,
\end{equation}
which yields 
\begin{equation}
\Im C^-(\Omega)=-\frac{1}{2}J_\nd^2\sum_{\alpha,\s=\pm}\s\mc{L}_2(\Omega+\alpha V+\s h)M\,.
\end{equation}
For the imaginary part of the ac conductance~\eqref{eq:ImG} we finally obtain
\begin{equation}
\begin{aligned}[b]
&\Im G(\Omega)\\
&=-\frac{M}{2\Omega}J_\nd^2\!\!\sum_{\alpha,\s=\pm} \s\Big[\mc{L}_2(\Omega+\alpha V+\s h)-\mc{L}_2(\alpha V+\s h)\Big]\,.
\end{aligned}
\label{eq:ImG_result}
\end{equation}

The real part of the ac conductance \eqref{eq:ReG_final} and its imaginary part can be combined to a single complex function. Introducing $\mc{H}_2(x)=\mc{L}_2(x)+i\pi|x|_2/2$ the ac conductance is given by
\begin{equation}
\begin{aligned}[b]
G(\Omega)=\frac{3\pi}{4}J_\nd^2-i\frac{M}{2\Omega}J_\nd^2\!\!\sum_{\alpha,\s=\pm}\!\!\!\s\Big[&\mc{H}_2(\Omega+\alpha V+\s h)\\
&-\mc{H}_2(\alpha V+\s h)\Big]\,.
\end{aligned}
\end{equation}

\subsection{Low frequency limit}\label{ssec:LFL}

In this appendix we discuss the additional contributions to the current noise and the ac conductance arising from the reducible part of the current-current correlation function~\eqref{eq:C} 
\begin{equation}
C_{\red}^\pm(\Omega)=-i\Tr_\D\left[\Sigma_I(\Omega)\frac{1}{\Omega-L_\D^\eff(\Omega)}\Sigma_I^\pm(\Omega,i0^+)\rho_\D^\st\right]
\label{eq:Cred}
\end{equation}
in the low frequency limit. For $\Omega\rightarrow0$, the contribution of the eigenvalue $z_1$ appears to be of order $J^2$. In addition, the zero eigenvalue leads to a singularity. In the following we analyze these contributions.

First we compute the current kernels $\Sigma_I$ and $\Sigma_I^\pm$.
The kernel of the normal current operator $\Sigma_I$ derived in Ref.~\onlinecite{PhysRevB.80.045117} for zero frequency is obtained by replacing the Laplace variable by the frequency $\Omega$ 
\begin{equation}
\Sigma_I(\Omega)=i\frac{3\pi}{8}VJ_\nd^2L^b+\frac{1}{8}J_\nd^2\!\sum_{\alpha,\s=\pm}\!\!\!\alpha\s \mc{H}_2(\Omega+\alpha V+\s h)L^{1z}\,.
\label{eq:SigmaI}
\end{equation}
For the derivation of the kernel $\Sigma_I^\pm(\Omega,\xi)$ we have to take into account all diagrams withe the current vertex on arbitrary position. In particular, we have to distinguish between the two frequencies, $\Omega$ occurring to the left of $I^\pm$, and $\xi\rightarrow0^+$ to the right. Furthermore, since $\Sigma_I(\Omega)$ is on the leftmost position in Eq.~\eqref{eq:Cred} we can neglect all contributions vanishing under the trace. Differently for $\Sigma_I^\pm(\Omega,\xi)$ all terms have to be taken into account. 
This yields the RG equation 
\begin{equation}
\begin{aligned}[b]
&\ddl{\Sigma_I^\pm(\Omega,\xi)}=-i\bI^\pm_{12}(\Omega,\xi)K(\xi_{12})\bG_{\2\1}(\xi_{12})\\
&\qquad-i\bG_{12}(\Omega)K(\Omega_{12})\bI^\pm_{\2\1}(\Omega_{12},\xi)\\
&\qquad-2i\bG_{12}(\Omega)K(\Omega_{12})\bI^\pm_{\23}(\Omega_{12},\xi)K(\xi_{\23})\bG_{\3\1}(\xi_{\23})\\
&\qquad-2i\bG_{12}(\Omega)K(\Omega_{12})\bG_{\23}(\Omega_{12})K(\Omega_{13})\bI^\pm_{\3\1}(\Omega_{13},\xi)\\
&\qquad-2i\bI^\pm_{12}(\Omega,\xi)K(\xi_{12})\bG_{\23}(\xi_{12})K(\xi_{13})\bG_{\3\1}(\xi_{13})\,.
\end{aligned}
\label{eq:RGeq_SigmaIpm}
\end{equation}
Solving this equation in an analog way as the one for $\Sigma_{\II}^\pm$, we obtain
\begin{subequations}
\label{eq:SigmaIpm}
\begin{align}
\label{eq:SigmaIp}
&\begin{aligned}[b]
\Sigma_I^+&(\Omega,0^+)=-\frac{1}{2}J_\nd^2\sum_{\alpha,\s=\pm}\alpha \s\mc{H}_2(\Omega+\alpha V+\s h)L^{3z}\\
&+i\frac{\pi}{2}J_\nd^2V(3L^b-L^a)+i\frac{\pi}{2}J_\nd^2\sum_{\s=\pm}\s |V+\s h|_2L^{1z}
\end{aligned}\\
\label{eq:SigmaIm}
&\begin{aligned}
\Sigma_I^-(\Omega,0^+)=\;&2J_\nd^2\sum_{\s=\pm}\mc{L}_2(V+\s h)L^a\\
&-J_\nd^2\sum_{\alpha,\s=\pm}\alpha\mc{H}_2(\Omega+\alpha V+\s h)L^a\,.
\end{aligned}
\end{align}
\end{subequations}

Now we discuss the contribution of the eigenvalue $z_1$.
With \eqref{eq:SigmaI} and \eqref{eq:SigmaIp}, the reducible contribution at $\Omega=0$ to the symmetric current-current correlation function is given by
\begin{align}
C_\red^+(0)|_{z_1}&=\Tr_\D\Sigma_I(0)\frac{1}{\Gamma_1}P_1\Sigma_I^+(0,i0^+)\rho_\D^\st\nonumber\\
&=-2\pi J_\nd^4\frac{1}{\Gamma_1}VM\sum_{\s=\pm}\s|V+\s h|_2\nonumber\\
&\quad-\left(2M^2+\frac{1}{2}\right)J_\nd^4\frac{1}{\Gamma_1}\left[\sum_{\s=\pm}\s|V+\s h|_2\right]^2\,.
\label{eq:Sp_Omega0}
\end{align}
Since $C_\red^+(0)|_{z_1}$ is real and $\Re C^+=S^+$, it represents the reducible contribution to the symmetric noise.
Since $\Sigma_I^-(0,0^+)=0$ there is no additional contribution to the antisymmetric noise at $\Omega=0$.

In the limit $\Omega\rightarrow0$, Eq.~\eqref{eq:G} for the ac conductance reads 
\begin{equation}
G(\Omega\rightarrow0)=\dd{C^-}{\Omega}\bigg\vert_{\Omega=0}\,.
\end{equation}
Using Eq.~\eqref{eq:Cred} the reducible part of the conductance can be expressed directly by the current kernels $\Sigma_I$ and $\Sigma_I^-$
\begin{equation}
G_\red(\Omega\rightarrow0)|_{z_1}=-\frac{1}{\Gamma_1}\Tr_\D\left[\Sigma_I(0)P_1\dd{\Sigma_I^-}{\Omega}\bigg\vert_{\Omega=0}\rho_\D^\st\right]\,,
\end{equation}
where we used that $P_1$ and $\Gamma_1$ are independent of the external frequency $\Omega$ and $\Sigma_I^-(0,0)=0$. With \begin{equation}
\dd{\Sigma_I^-}{\Omega}\bigg\vert_{\Omega=0}=-2i\dd{\Gamma_1}{V}L^a
\end{equation}
we obtain
\begin{equation}
G_\red(\Omega=0)|_{z_1}=-\frac{\pi}{2}J_\nd^2\frac{M}{\Gamma_1}\dd{\Gamma_1}{V}\sum_{\s=\pm}\s|V+\s h|_2\,.
\label{eq:G_Omega0}
\end{equation}
We finally note that this term is real and thus does not affect the imaginary part of the ac conductance.

The reducible contribution of~\eqref{eq:Cred} of the zero eigenvalue is given by
\begin{equation}
C_\red^\pm(\Omega)|_{z_0}=-i\Tr_\D\Sigma_I(\Omega)\frac{1}{\Omega}P_0\Sigma_I^\pm(\Omega,0^+)\rho_\D^\st\,,
\end{equation}
which leads to a singularity in the limit $\Omega\rightarrow0$.
To study the singular behavior in detail we use the kernels $\Sigma_I(\Omega)$ and $\Sigma_I^\pm(\Omega,0^+)$ given by~\eqref{eq:SigmaI} and \eqref{eq:SigmaIpm} respectively. The contribution to $S^+(0)$ is 
\begin{equation}
\begin{aligned}[b]
S_\red^+(\Omega)|_{z_0}&=\Re C_\red^+(\Omega)|_{z_0}\\
&=-\pi\Tr_\D\Sigma_I(\Omega)P_0\Sigma_I^+(\Omega,0^+)\rho_\D^\st\delta(\Omega)\,,
\end{aligned}
\end{equation}
due to $\Im\frac{1}{\Omega+i\delta}=-\pi\delta(\Omega)$.
Inserting \eqref{eq:SigmaI} and \eqref{eq:SigmaIp} we find
\begin{align}
S^+_\red(0)|_{z_0}&=\begin{aligned}[t]\bigg[&\frac{9\pi^3}{8}J_\nd^4V^2+\frac{3\pi^3}{2}J_\nd^4VM\sum_{\s=\pm}|V+\s h|_2\\
&+\pi^2J_\nd^4\Big(M\sum_{\s=\pm}|V+\s h|_2\Big)^2\bigg]\delta(\Omega)
\end{aligned}\nonumber\\
&=2\pi\Ew{I}_\st^2\delta(\Omega)\,,
\end{align}
where we used~\cite{PhysRevB.80.045117}
\begin{equation}
\Ew{I}_\st=\frac{3\pi}{4}J_\nd^2V+\frac{\pi}{2}J_\nd^2M\sum_{\s=\pm}\s|V+\s h|_2\,.
\label{eq:Ist}
\end{equation}
Thus this term is exactly canceled by the second term of Eq.~\eqref{eq:S_Ca} and does not lead to any singularities.

As $\Sigma_I^-\propto L^a$ and $P_0 L^a=0$ there is no singular behavior in $S^-(0)$ and hence $G(\Omega\rightarrow0)$ is not affected.


\begin{thebibliography}{}
\bibitem{note1} Derivatives are performed with respect to the explicit $\Omega$ or $V$
dependence in the magnetization or in $|.|_i$ only, as the derivative of $J(\Lambda_c)$, 
the decay rates, or the $\arctan$-functions would introduce higher-order corrections not 
treated consistently.
\bibitem{note2} To capture the variation of the additional term at small  $\Omega$ we replaced 
$1/\Gamma_1$ by $i/(\Omega+i\Gamma_1)$.
\end{thebibliography}
\end{document}